\documentclass[reprint,prl,aps,showpacs,twocolumn,superscriptaddress]{revtex4-2} 

\usepackage{graphicx}
\usepackage{epsf}
\usepackage{epstopdf}
\usepackage{cancel}
\usepackage{color}
\usepackage{float}
\usepackage{soul} 
 \usepackage{amsmath,amssymb}

\newcommand*{\BeginNoToc}{%
	\addtocontents{toc}{%
		\edef\protect\SavedTocDepth{\protect\the\protect\value{tocdepth}}%
	}%
	\addtocontents{toc}{%
		\protect\setcounter{tocdepth}{-10}%
	}%
}

\begin{document}

\title{Physics of selective conduction and point mutation in biological ion channels}

\author{W.A.T.~Gibby}
	\email{w.gibby@lancaster.ac.uk}
	\affiliation{Department of Physics, Lancaster University, Lancaster LA1 4YB, UK.}
\author{M.L.~Barabash}
	\affiliation{Department of Physics, Lancaster University, Lancaster LA1 4YB, UK.}
    \author{C.~Guardiani}
	\affiliation{Department of Physics, Lancaster University, Lancaster LA1 4YB, UK.}
	\affiliation{Department of Mechanical and Aerospace Engineering, Sapienza University, Rome, Italy}
\author{D.~G.~Luchinsky}
    \affiliation{Department of Physics, Lancaster University, Lancaster LA1 4YB, UK.}
    \affiliation{KBR, Inc., Ames Research Center, Moffett Field, CA, USA}
\author{P.V.E.~McClintock}
	\email{p.v.e.mcclintock@lancaster.ac.uk}
    \affiliation{Department of Physics, Lancaster University, Lancaster LA1 4YB, UK.}

\date{\today}

\begin{abstract}   
We introduce a statistical and linear response theory of selective conduction in biological ion channels with multiple binding sites and possible point mutations. We derive an effective grand-canonical ensemble and generalised Einstein relations for the selectivity filter, assuming strongly coordinated ionic motion, and allowing for ionic Coulomb blockade. The theory agrees well with data from the KcsA K$^+$ channel and a mutant. We show that the Eisenman relations for thermodynamic selectivity follow from the condition for fast conduction and find that maximum conduction requires the binding sites to be nearly identical.

\end{abstract}

\maketitle

Understanding, predicting and optimising the ionic transport properties of nanopores remains a critical challenge to both nanotechnology \cite{wang2017fundamental} and biophysics \cite{Hille:01,roux2017ion}. Interest is motivated in large part by the importance and diversity of the applications, which include water purification \cite{mauter2018role}, DNA sequencing \cite{deamer2016three}, and biological ion channels together with their role in medicine \cite{Hille:01,Ashcroft1999,Huang2014}.

\begin{figure}[t!]
	\includegraphics[width=1\linewidth,trim={0 0 0cm 0},clip]{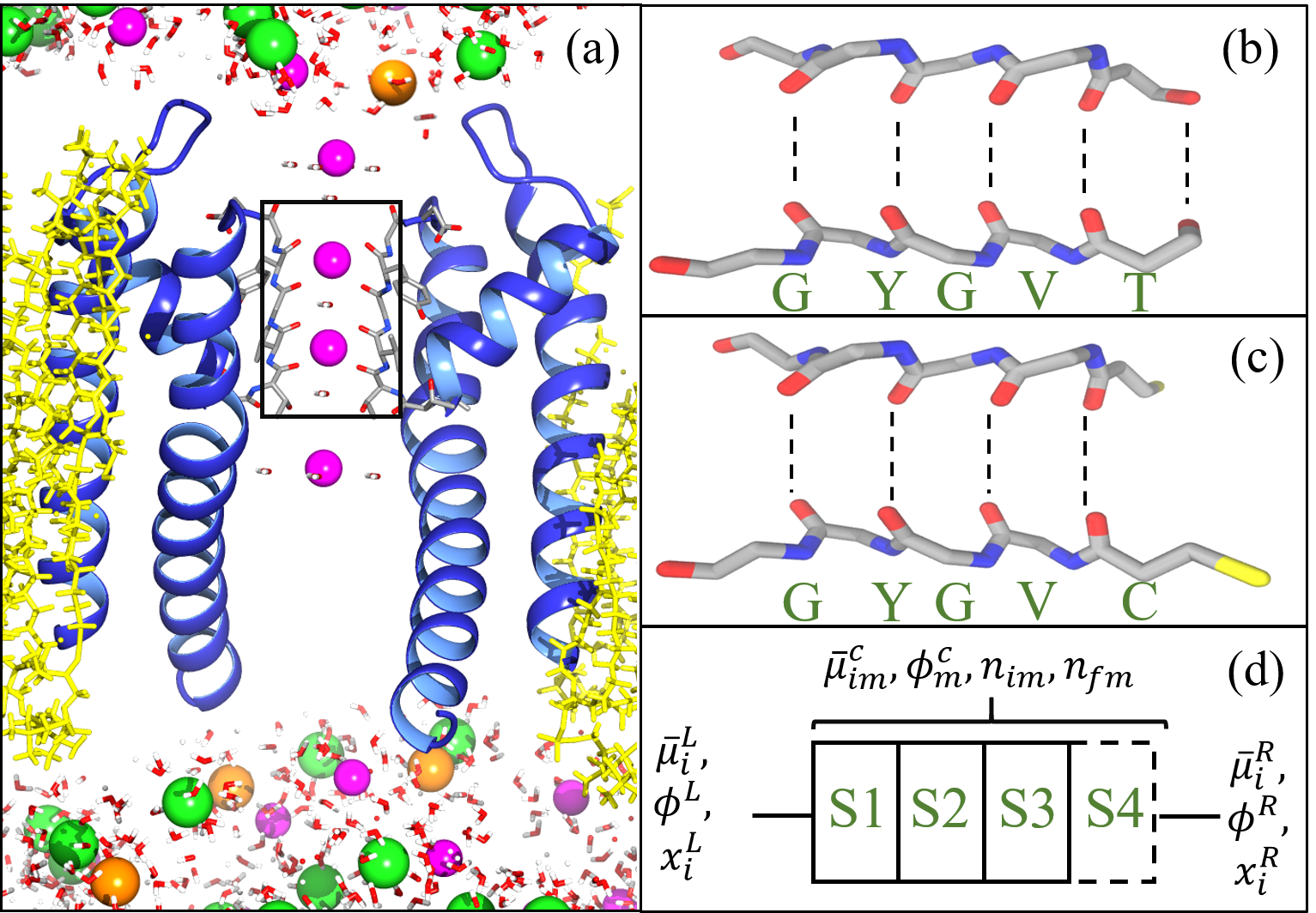}
	\caption{\label{fig:KcsAscheme} Structure of open KcsA (5vk6.pdb) \cite{cuello2017gating} visualised using chimera \cite{pettersen2004ucsf}. (a) Two chains (blue ribbons) spanning a lipid membrane (yellow strands) between two aqueous ionic solutions.  The SF is located within the box, and K$^+$ (purple), Na$^+$ (brown), and Cl$^-$ (green) ions alongside water molecules are included. (b) Structure of the SF for wild type KcsA and (c) the T75C mutant, with indicated amino acids at binding sites. (d) Lattice model used to define the system.  }
\end{figure}

Biological channels are proteins with a central pathway (nanopore), spanning a lipid membrane. Their primary function of selectively conducting ions at nearly the diffusion rate is effected mainly by a narrow selectivity filter (SF) (Fig.~\ref{fig:KcsAscheme}). Key structural features of K$^+$ SFs, elucidated over decades~\cite{Hille:01,Aidley:00,Chung:07}, include a sequence of sub-nanometer sized binding sites that can have strongly charged residues whose affinities are site and ion-specific.

It is found that point mutations can greatly influence permeation and selectivity~\cite{Heginbotham1994,zhou:2004,Derebe2011b,oakes2019insights,Tilegenova2019,labro2018inverted,oakes2019insights} by modifying the affinities, geometry and residual charge at individual sites. Predicting these changes is a long-standing and challenging problem. Nano-confinement of ions in the SF is affected by e.g.\ partial charges \cite{bucher2006polarization,kraszewski2009determination}, ionic diffusivity \cite{Tieleman:01}, electrical permittivity \cite{schutz2001dielectric,ng2008estimating}, quantum mechanical interactions \cite{varma2010multibody,varma2011design,Bucher:10,de2018roles}, and the species-dependent positions of binding sites \cite{thompson2009mechanism,Nimigean2011b,Kim:11,de2018roles}.


Particularly intriguing problems associated with point mutations are understanding the relations between selectivity, conductivity \cite{nimigean2002na,Nimigean2011b,thompson2009mechanism,medovoy2016multi,kim2011selective}, highly coordinated conduction mechanism \cite{Doyle:98a, MacKinnon2001b, Koepfer2014, kopec2018direct,barabash2019potential,liu2013collective,sumikama2019queueing,coates2020ion}, and occupancy of individual sites and the SF as a whole.  It is clear that valuable insight can be gained from a fundamental theory.

Earlier statistical theories were focused primarily either on the problem of selectivity, or on that of conductivity. Thermodynamic selectivity \cite{Eisenman1983} in this context is defined as the difference in interaction energy in the bulk and in the channel, which differs between species. In K$^+$ channels it led to the snug-fit model highlighting the importance of close coordination of ions by charged oxygen atoms~\cite{bezanilla1972negative,Doyle:98a}; and it has been analysed at the scale of individual binding sites in many channel types see e.g.\ \cite{gillespie2002physical,Gillespie2002a,gillespie2003density,nonner2008volume,gillespie2014selecting}.  Ionic conduction occurs via a knock-on mechanism \cite{Hodgkin:55,MacKinnon2001b,Berneche:01,Zhou2003}, which has been investigated using statistical physics~\cite{Yesylevskyy:05,Kharkyanen:10}. This approach led to an important analogy between ionic Coulomb blockade (ICB) and electronic Coulomb blockade in quantum dots \cite{Krems:13,Kaufman:13a,Kaufman:15}, which also served to highlight the importance of long-range interactions for valence selectivity  \cite{Shklovskii:05,Shklovskii:06,Krems:13,Kaufman:13a,Kaufman:15,Feng:16,kavokine2019ionic}. Furthermore, statistical and information theories have provided insight into the binding of ions and its relationship with the potential of mean force \cite{Roux1999,Im:00,Roux:04,rogers2011information}.

These insightful theories have often ignored, however, the multi-component and multi-site nature of biological SFs and do not account for the ion-specific affinities of individual binding sites and the fact that these can be altered by point mutations. A theory able to encompass these phenomena is crucial for understanding the properties of real SF's and might also shed light on phenomena such as ICB~\cite{Feng:16,kavokine2019ionic}, anomalous mole fraction effect~\cite{Kaufman:17a,Nonner:98b,gibby2019theory}, and the mechanism of knock-on~\cite{kopec2018direct,Tilegenova2019}, all of which are subjects of extensive debate.

In this Letter we introduce such a theory, based on statistical physics and linear response. It relates both the kinetic and thermodynamic properties of permeation directly to pore structure, and it shows that Eisenman's selectivity relation follows from the condition for maximum conductivity of the SF. Furthermore, it allows one to calculate optimal transport parameters, and the conductivities of individual sites and of the SF as a whole. It opens the way to statistical analysis of point mutations in biological channels, including mutation-induced changes in conduction, selectivity, and population.

The K$^+$-conducting channel KcsA is shown in Fig.~\ref{fig:KcsAscheme}. Its pore $(c)$ has average radius $R_c \sim$2\AA, length $L_c\sim$12\AA \;, volume $V_c$, and 4 binding sites formed by charged oxygen atoms in carbonyl/hydroxyl groups \cite{Doyle:98a,MacKinnon2001b}. The pore is thermally and diffusively coupled to the left ($ L $) and right ($ R $) bulk reservoirs $(b)$ (bottom and top respectively in Fig. 1(a)). Each bulk contains mixed solutions of $S$ total ionic species where $i\in 1,\cdots,S$. The primary function of the pore is conduction of K$^+$ at close to the rate of free diffusion whilst selecting strongly against Na$^+$.
The system as a whole is characterized by the canonical ensemble with constant total particle number $N_i$, volume $V$ and temperature $T$. Ions are free to leave the bulk solution and bind at specific sites in the pore, with $N^b_i$ being the total number of ions in either bulk prior to any binding.Due to the narrowness of the pore, ionic motion is confined to one dimensional conduction via  a finite number  of binding sites $M$, where each site $m \in 1,\cdots, M$ can hold a single ion at most. Therefore, the total number of ions in the pore is  $ n  = \sum_m^M\sum^S_in_{im} \leq M$, where  $\sum^S_{i=1}n_{im}\in 0,1$ and $n_i=\sum^M_{i=1}n_{im}$. All possible configurations
of ionic binding in the pore $\{n_{im}\}$ (or $\{n_{j}\}$ to simplify notations) are mutually exclusive leading to Fermi statistics~\cite{rogers2011information}. The energy of each configuration is found by explicitly counting the number of ions of each species that leave the left ($n''_{im}$) and right ($n'_{im}$) bulks and enter site $m$ in the pore (keeping the total number of ions constant)
\begin{align}
&E(\{n_j\}) =E_0+\mathcal{E}(\{n_j\})+\sum_{i=1}^S\left(N^L_i-\sum_{m=1}^M n''_{im}\right)\mu^L_i
\nonumber \\&
+\sum_{i=1}(N^R_i-\sum_{m=1}^M n'_{im})\mu^R_i +kT\ln(n_0)! +\sum^S_{i=1} kT\ln n_i!
\nonumber \\&
 +\sum_{i=1}^S\sum_{m=1}^M (n_{im}'+n_{im}'')(\bar\mu^c_{im}+qz_i\phi^c_m)
.\label{eq:Energy}
\end{align} \noindent Here we define  the thermodynamic part of the energy $E_0 = TS-pV$, entropy $S$, pressure $p$, the long range interaction energy $\mathcal{E}$ between ions and fixed charges, and $n_0$ is the number of empty sites (that may be occupied by water molecules).

It can be seen from \eqref{eq:Energy} that the bulk electrolytes and binding sites of the pore represent a system with several interpentrating solutions, each characterized by its own chemical potential. The electrochemical potential in the bulk is defined~\cite{widom1982,Roux:04,Gillespie:08} as the sum of ideal, excess, and electrostatic parts
\begin{equation}\label{eq:chem-pot-b}
 \mu^b_i = kT\ln\left( \dfrac{(\Lambda^{b}_i)^3 x^b_i}{q^{\text{int},b}_i}\right) + \bar{\mu}^b_i + qz_i\phi^b,
\end{equation}
where $\Lambda^b_i$, $q^{\text{int},b}_i$, $x^b_i$, $z_i$, $\phi^b$ and $\bar{\mu}^b_i$ denote the thermal wavelength, internal partition function, mole fraction, valence,  external electric and excess chemical potential respectively.

For a binding site, the  electrochemical potential is \cite{Kirkwood1935}
\begin{equation}\label{eq:chem-pot-c}
	\mu^c_{im} = kT\ln \dfrac{(\Lambda^{c}_i)^3 f\left(\{n_j\}\right)}{q^{\text{int},c}_i} + \bar{\mu}^c_{im} + qz_i\phi^c_m + \Delta\mathcal{E}. 
\end{equation}
It is characterized (see SI) by values of the excess chemical $\bar{\mu}^c_{im}$ and electrostatic $qz_i\phi^c_m$ potentials at each site, the change in the energy of interaction between ions  $\Delta \mathcal{E}$ when one ion is added to the pore, and the factor $f$ accounting for indistinguishably of ions in the pore. We assume that $(\Lambda^{b}_i)^3/q^{\text{int},b}_i\sim (\Lambda^{c}_i)^3/q^{\text{int},c}_i$ and so it has been factored out of our expression \eqref{eq:Energy} for the total energy.

We note that ion transition from the bulk to the pore results in small fluctuations of the total energy in \eqref{eq:Energy}. This allows us to derive~\cite{Landau1980,rogers2011information} the effective \textit{grand canonical ensemble} (GCE) for the pore by factorizing the partition function as a product of bulk and pore constituents and cancelling constant terms
\begin{align}
	&	P(\{ n_j \}) = \mathcal{Z}^{-1} \left( \frac{1}{n_0!}\prod_i \frac{1}{n_i!}\right) \nonumber \\ &
		\times
	\exp\left[\left(\sum_{i=1}^S\sum_{m=1}^M n_{im}\Delta \tilde\mu^b_{im}-\mathcal{E}(\{n_j\})\right)/kT\right]\label{eq:StatIndisGCEProb}
\end{align}
where $\Delta \tilde{\mu}^b_{im}= \Delta \bar\mu_{im}^b+qz_i\Delta \phi_m^b +kT\log(x_i^b) $ and the partition function $\mathcal{Z}$ represents the normalisation enforcing $\sum_{\{n_j\}} P\equiv 1$. Note that $\Delta$ is defined by the difference between pore and bulk, so $\Delta \bar\mu_{im}^b =\bar\mu_i^b-\bar\mu^c_{im} $ etc.

The corresponding free energy ($G=E-TS+pV$) is,\begin{align}
	&	G(\{n_j\}) =\mathcal{E}(\{n_j\})   +kT\ln n_0!
		\nonumber \\
	&   + \sum^S_{i=1} kT\ln n_i!-\sum_{i=1}^S\sum_{m=1}^M n_{im}[\Delta \bar\mu^b_{im}+kT\ln(x^b_i)].\label{eq:FREE}
\end{align}
The derived model is consistent with many earlier theoretical results~\cite{Chesnut1963,Roux1999,Roux:04,rogers2011information,hughes2014introduction} and accounts for the key features of selective conduction in biological SF's, including ionic Coulomb blockade and structure of the individual sites in the SF.

The grand potential ($\Omega = -kT \ln \mathcal{Z}$) can now be used to compute the occupancies
of the sites and the SF
\begin{align}\label{eq:occupancy}
    \langle n_{im} \rangle   = \sum_{\{n_j\}}n_{im} P(\{n_j\})
    \qquad \langle n_i \rangle
        = \sum_m\langle  n_{im} \rangle.
\end{align}These are used to calculate the conductivity $\sigma_{im}$  (defined via the static density susceptibility $\chi_{im}$, see Eq.\ (25) of SI) at each site, and the total average conductivity $\sigma^T$ of the SF \begin{equation}\label{eq:sigmaT}
\sigma_{im}=z_iq^2\chi_{im}D_{im}., \quad \sigma_i^T=\left(\sum_m^M \sigma_{im}^{-1}\right)^{-1},
\end{equation}
where $D_{im}$ is the species diffusivity at site $m$.
It is clear that, in general, local geometry influences directly the conductivity at the site. For single file motion, conduction through the sites of length $L_{im}$ and cross-sectional area $A_{im}$ connected in series (see Fig.~\ref{fig:KcsAscheme}) the conductance $\mathcal{G}^T$ and total current across the pore are\begin{equation}\label{eq:current}
     I = \mathcal{G}^T \mathcal{V}, \quad \mathcal{G}^T=\sum_i^S\left(\sum_m^M\frac{L_{im}}{A_{im} \sigma_{im}}\right)^{-1}.
\end{equation}Equations \eqref{eq:occupancy}-\eqref{eq:current} allows us to calculate occupancy, conductivity, and current for each binding site, and for the SF as a whole, for any given mutation of the structure. To compare these results to experimental data, we consider conduction of the wild-type (WT) KcsA filter and its mutant T75C (MuT) obtained by point mutation~\cite{zhou:2004} of site S4 of KcsA, see Fig.~\ref{fig:KcsAscheme}.

We assume that: bulk solutions contain Na$^+$ and K$^+$ at concentrations of 0.2M; ions may occupy neighbouring sites; and the occupancy of the SF is restricted to 3 ions. Under these plausible conditions there are 65 configurational states in the system (see SI). To estimate the energies $G(\{n_i\})$ of these states we use Eqs. \eqref{eq:StatIndisGCEProb} and \eqref{eq:FREE} and approximate the total electrostatic energy of the pore $\mathcal{E}$ as a capacitor~\cite{Krems:13,Kaufman:15} of capacitance $C$, total charge $Q = \left(n_f + \sum_i n_i\right)q$, and charging energy $U_c\sim 10kT$\begin{equation}\label{eq:electr}
    \mathcal{E} = U_c\left[\sum_m \left(\sum_i n_{im}+n_{fm}\right)\right]^2; \quad  U_c= \frac{Q^2}{2C}.
\end{equation}Here $n_{fm}$ is an effective valence of the binding sites, $n_f = \sum_m n_{fm}$, and  $q$ is the unit charge. There are alternative approximations including~\cite{Kitzing1992,Shklovskii:06,Yesylevskyy:05,Krems:13,Kaufman:15}. However, without experimental evidence to the contrary, \eqref{eq:electr} provides a simple and physically appealing interpretation of $\mathcal{E}$ by analogy with quantum dots~\cite{Beenakker:91}.

\begin{figure}[t!]
 	\centering
	\includegraphics[scale=0.35]{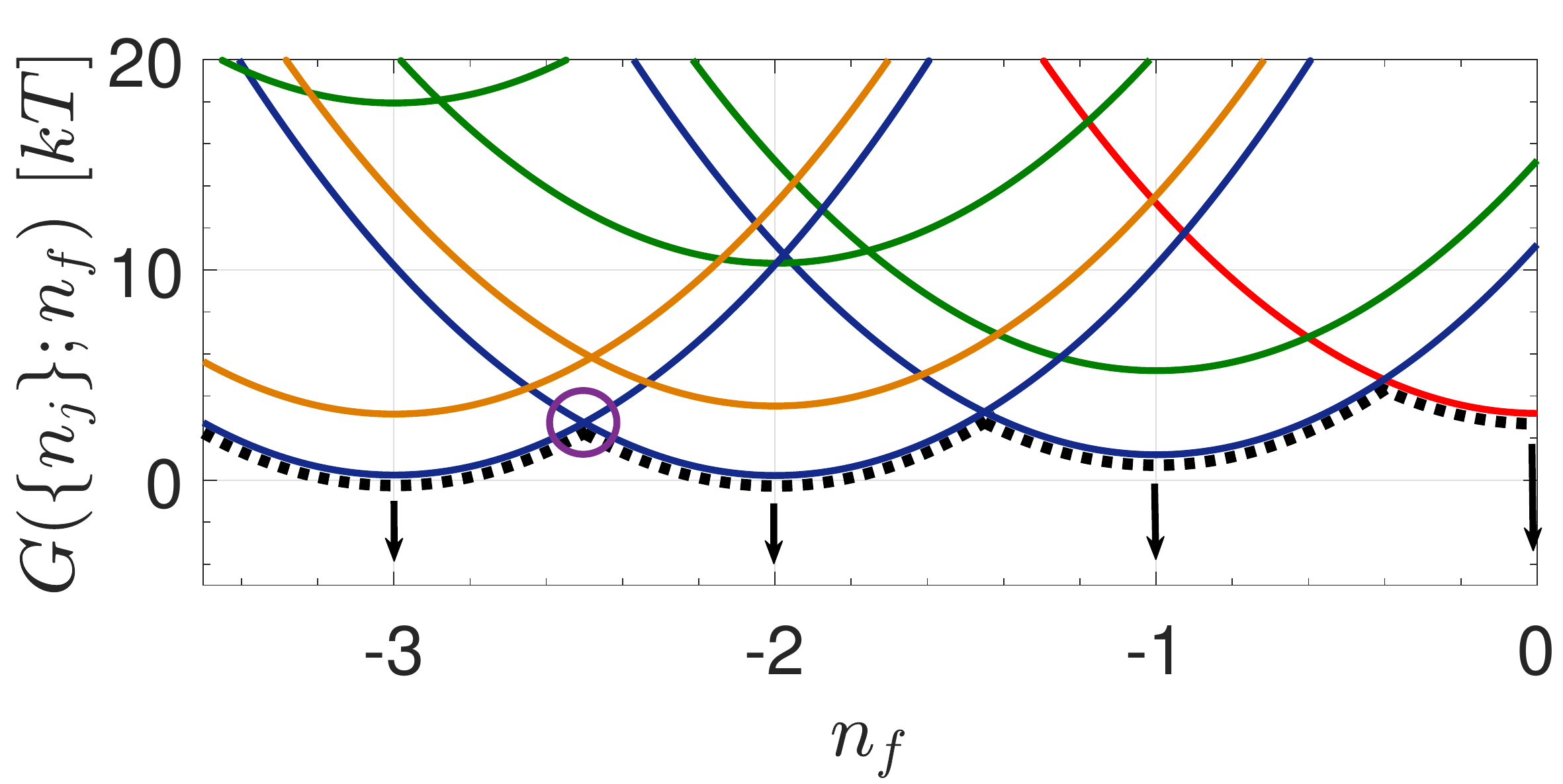} \\
 	\caption{\label{fig:EnergySpectrums2D}  Free energy \textit{vs.} $n_f$ for WT with KCl and NaCl solutions at $0.2$M. Only the most favoured states are included. Colours, red, blue, green, orange and black dashed denote empty, pure K$^+$ and Na$^+$, mixed  and the ground states respectively. The purple circle shows when  the pore is equally energetically favourable to hold 2 or 3 K$^+$ ions. $\Delta\bar\mu_{K,1-4}$  in the WT are $\sim 6.2, 5.7, 6, 6.2 kT$ while $\Delta\bar\mu_{Na,1-4}$ are $\sim 2.2, -2.6, -1.6, 0.1$. }
\end{figure}

A subset of the calculated lowest energy levels is shown in Fig. \ref{fig:EnergySpectrums2D} as a function of $n_f$. Each curve is parabolic and has a minimum when the total charge within the pore is neutralised ($Q=0$). These minima correspond to {\it non-conducting} ground states of the SF that hold 0, 1, 2, or 3 potassium ions as shown by arrows.

According to ICB theory, {\it conducting} states correspond to the degeneracies where the lowest energy levels intersect, cf.~\cite{Yesylevskyy:05}. An example of this situation where 2-ion and 3-ion K$^+$ states are degenerate is highlighted by the purple circle. In accordance with MacKinnon's idea of charge balance \cite{Zhou2003}, the total fixed charge is close to $Q_f=-2.5q$ because this corresponds to an average of 2.5 ions inside the pore. If we assume that all oxygen atoms are equally partially charged then we estimate their individual valences to be $z_0 \sim 0.125$ and find that the total charge on an 8-oxygen-caged binding site is $-1q$.

\begin{figure}[t!]
 	\centering
	\includegraphics[width=\linewidth]{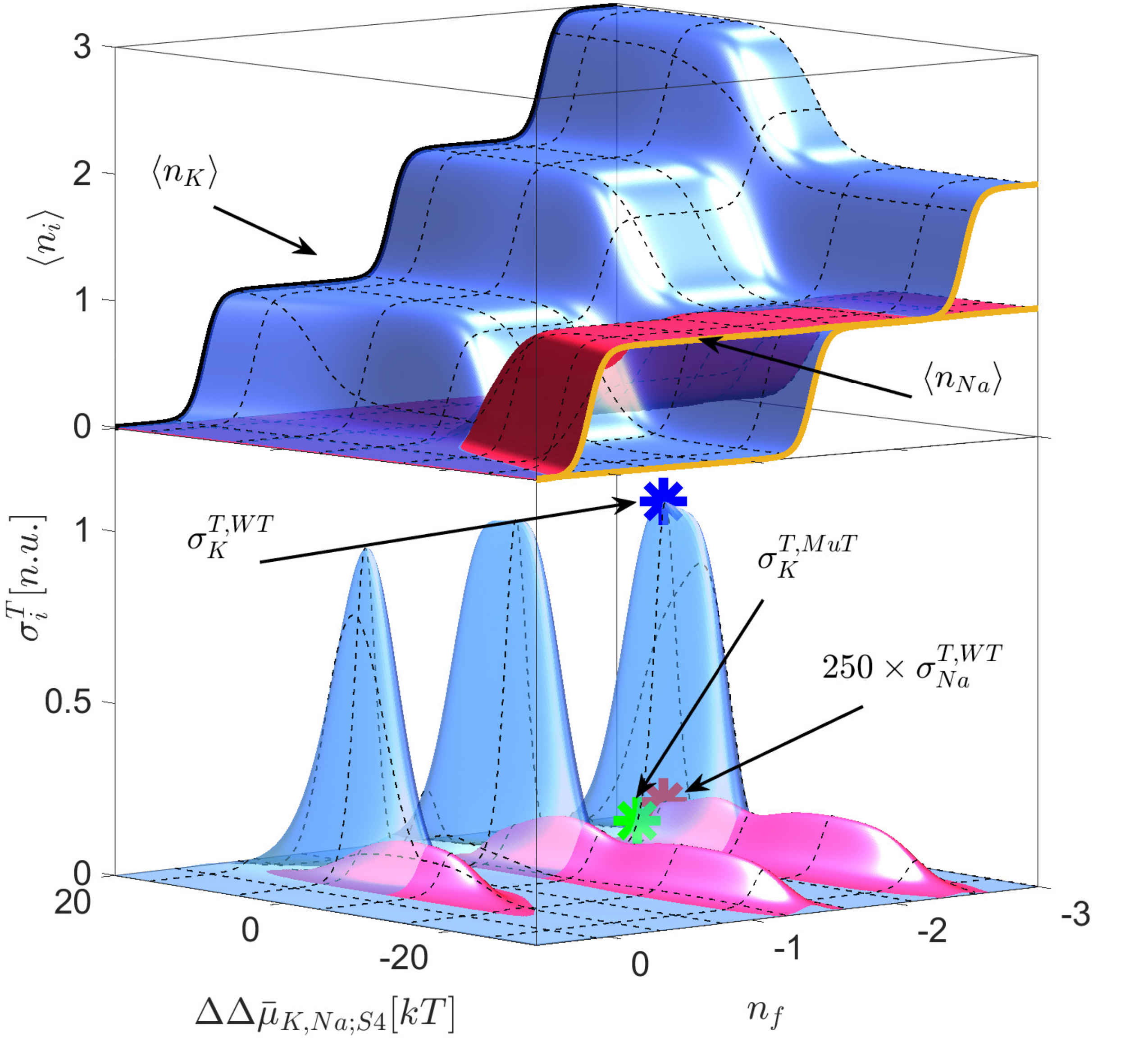}
	\caption{\label{fig:total}  K$^+$ (blue) and Na$^+$ (red) occupancy (top) and conductivity (bottom) $vs.$  $n_f$ and $\Delta\bar\mu_{K,Na;S4}$ in symmetrical 0.2M mixed bulk solutions. Conductivity and occupancy form a set of resonant peaks and steps respectively. Each step maximises under the conditions of barrier-less knock on, it being the favoured species and minimal difference in site affinity.  Using identical  parameters to Fig.~\ref{fig:EnergySpectrums2D} obtained through experimental comparison, we indicate the WT K$^+$ (blue) and Na$^+$ (red) and MuT   K$^+$ (green) conductivity's via coloured stars. Selectivity appears via the shift in both occupancy and conductivity from K$^+$ (blue) to Na$^+$ (red) surfaces, and the conductivity ratio yields $\sigma_{K}^{T,WT}/\sigma^{T,WT}_{Na}\sim 2 \times 10^3$.  }
\end{figure}

The vertical shifts of the levels are determined by the values $\Delta\bar\mu_{im}$. These parameters are extracted through comparison with experimental data \cite{zhou:2004} and results of molecular dynamic simulations~\cite{Kim:11}, including of the site's occupancies and current-voltage relations, see below. Once the energy levels of the system states are known, we can calculate the occupancy and conductivity of each binding site and of the SF as a whole using Eqs.\ \eqref{eq:StatIndisGCEProb}-\eqref{eq:sigmaT}. Calculations of the multi-component occupancy $\left< n_i \right>$ and conductivity $\sigma_i^T $ of the SF in the presence of mixed KCl and NaCl solutions are shown in Fig.~\ref{fig:total}.

In general,  $\left< n_i \right>$ and $\sigma_i^T $ are complex multi-parametric surfaces. Here we plot their dependence on the selectivity (and thus affinity) of site S4 and the total wall charge $n_f$. Note the following key features of these plots. First, that Eqs.\ \eqref{eq:sigmaT} and \eqref{eq:current} take explicit account both highly coordinated motion of ions in the channel (see Eqs. (8)-(10) of SI) and the conductivity of individual sites in the presence of long range interactions. Secondly, the conductivity of the SF is smaller than the smallest site conductivity. Hence optimized conductance of the SF corresponds to nearly identical binding sites in line with experimental results, see Fig.~\ref{fig:KcsAscheme}. Thirdly, the whole SF becomes non-conducting when one site ceases to conduct. The SF conductivity resonates strongly as a function of both wall charge and $\Delta\bar\mu_{K,4}$. Therefore, a small change of parameters at a given site can inactivate the whole SF, thus illuminating a possible mechanism of C-type inactivation~\cite{Xu2019}.

The sensitive dependence of $\sigma_T$ on its parameters suggests that the KcsA SF must be carefully tuned to achieve fast, strongly selective, diffusion of potassium ions. The corresponding optimal parameters can now be found analytically. As mentioned above, maximum conduction occurs when the sites share affinity and lowest energy levels intersect: $G(n_K+1,n_f) - G(n_K,n_f) \sim 0$, which is equivalent to equilibrium between the bulks and the SF i.e.\ $\mu^b_i=\mu^c_i$ cf~\cite{Yesylevskyy:05}. Note that we neglect a small entropy contribution from the fact that sites are now identical (see SI). It follows directly that maximal conductivity occurs when,
\begin{equation}\label{OptimalMuBar}
    \bar\mu_{Km}^{c,*}=  \bar\mu_K^b+kT\ln(x_K)-\Delta\mathcal{E}-\Delta f,
\end{equation}
 Eq.\ \eqref{OptimalMuBar} can be inverted to identify the optimal fixed charge for the SF. If we consider a pore with $n_f\sim -2.5$, and 0.2M solutions then we estimate $\Delta\bar\mu^{*}_{Km}\approx 6 kT$, consistent with experiment.

The conditions of maximal K$^+$ conduction and simultaneous strong selectivity ($\Delta G_{Na} = G(n_K+{Na}) - G(n_K+K)$) give us the free energy barrier for adding  Na$^+$ to site $m$ in the form
\begin{equation}\label{Eis}
    \Delta G_{Na} \sim  \Delta \tilde\mu_{K,m}-\Delta\tilde\mu_{Na,m} \gg kT,
\end{equation}
which is equivalent to the Eisenman selectivity relation~\cite{Eisenman1983}. Note that we have neglected the  entropic contributions due to mixing of ions and sites. Thus the theory resolves the long-standing conundrum~\cite{Roux:2014} of simultaneous fast conduction with strong selectivity of the KcsA SF, and shows that Eisenman's strong selectivity relation follows directly from the condition of fast conduction.

A numerical example illustrating this point is shown in Fig.~\ref{fig:total}. The state of the WT pore tuned for maximum conduction of K$^+$ ions and strong selectivity of K$^+$ over Na$^+$ is shown by the blue and red stars on the conductivity surfaces in the figure. The conductivity ratio  $\sigma_{K}^{T,WT}/\sigma^{T,WT}_{Na}\sim 2 \times 10^3$ is comparable with the commonly quoted ratio 1:1000 \cite{coates2020ion}.

Because different points of the multi-parametric conductivity and selectivity surfaces (Fig.~\ref{fig:total}) correspond to different experimental conditions (e.g.\ pH, concentrations) and mutations of the SF, the theory paves the way to detailed structure-function studies for many experimentally observed phenomena.

\begin{figure}[t!]
 	\centering
    \includegraphics[width=1\linewidth]{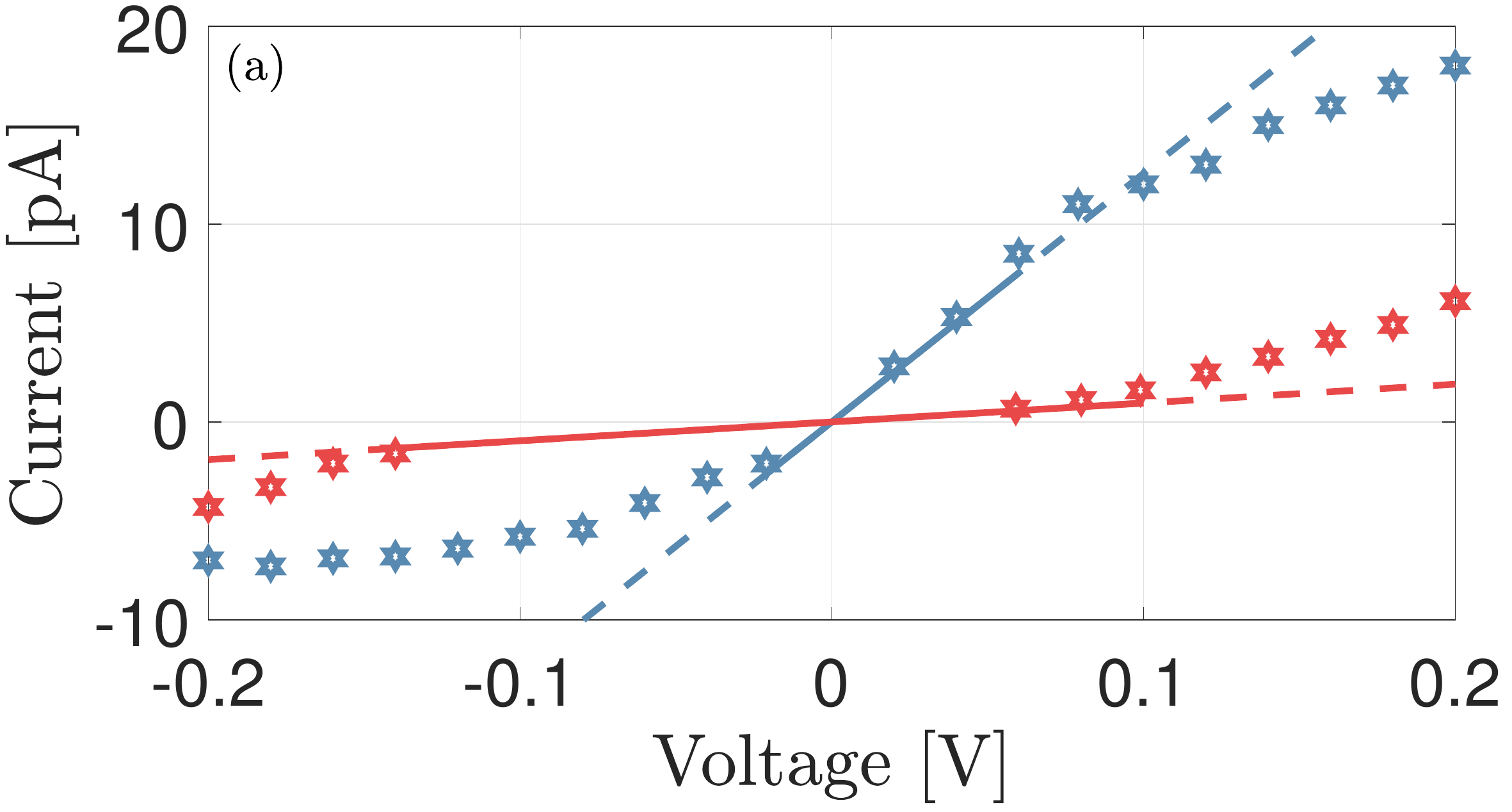}
    \includegraphics[width=0.45\linewidth]{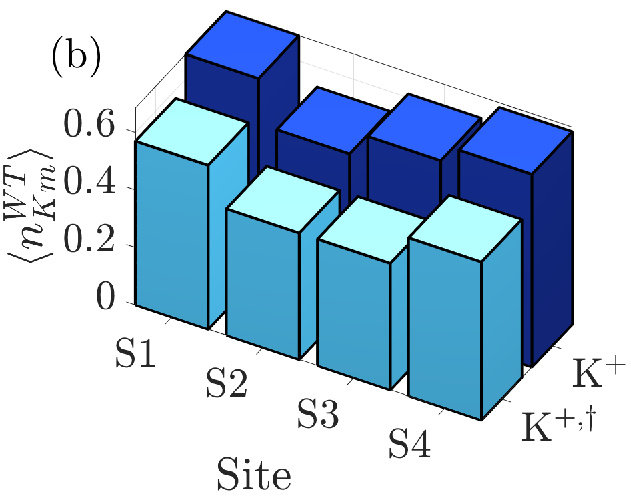}
	\includegraphics[width=0.45\linewidth]{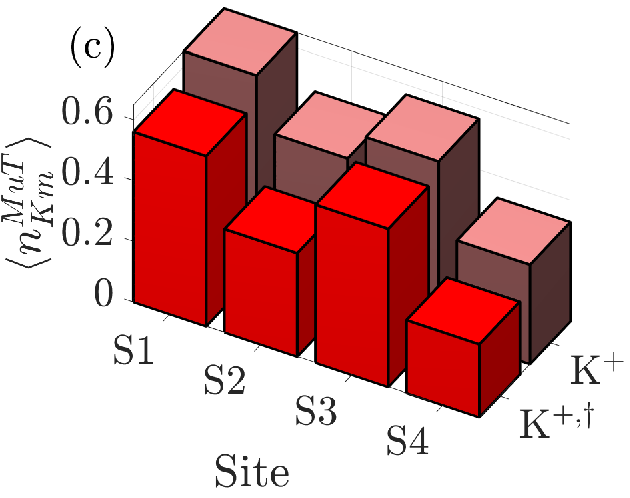}
	\caption{\label{fig:Fit}Comparison of theoretical current (a) (lines), and site occupancy (b) and (c) to experimental data \cite{zhou:2004} denoted by stars in (a) and $\dagger$ in (b) and (c). Theory and data are shown for the WT in blues and for the mutant (MuT) in reds. By comparison we find: $\Delta\bar\mu_{K,1-3}\sim 6.2,5.7,6 kT$, $\Delta\bar\mu_{K,4}^{WT}\sim 6.2 kT$ and  $\Delta\bar\mu_{K,4}^{MuT}\sim 5.2 kT$, a diffusivity in the channel of $2.3\times 10^{-10}$m$^2$s$^{-1}$; and finally, the new charge of the mutant was estimated to be $n_f\sim -2.32$.}
\end{figure}

Next we apply the theory to the analysis of the T75C point mutation in the KcsA SF~\cite{zhou:2004} replacing threonine with cysteine at location S4. This change does not significantly alter the side-chain volume but varies the electrostatic properties because the MuT lacks 4 hydroxyl ligands, effectively lowering the total attractive charge of the filter. Experiment demonstrates that the distribution of K$^+$ ions in the SF is modified between the WT and MuT and that it conducts potassium at a lower rate. WT and MuT are  highlighted by blue and red stars and K$^{+\dagger}$ bars in  Fig.~\ref{fig:Fit}.

To compare experimental results with theoretical predictions we take into account both the change in geometry and the fixed charge of the pore.  The modified state of the system is shown by green star in the Fig.~\ref{fig:total} and corresponds to the reduced charge of the SF from -2.5q to -2.32q, reduced  affinity of S4' from 6.2 to 5.2 $kT$,  and volume change of S4 by factor 1.2. The pore diffusivity in WT and MuT was estimated to be $3\times 10^{-10}$m$^2$s$^{-1}$, which is less than the bulk value, as expected \cite{Tieleman:01,Roux:04}. Using the $n_f$'s of the WT and MuT we can revise our earlier estimate, finding that the partial charge from each carbonyl group oxygen provides $-0.145q$ and the charge contribution from each hydroxyl group oxygens provide $-0.045q$.

The theoretical predictions are compared to the experimental data for current-voltage relation for the WT and MuT in Fig. \ref{fig:Fit} (a). The comparison extended beyond validity of the linear response regime is shown with dashed lines. The reduced conduction in the mutant due to the increased resistivity of the S4' site ($vs.$ S4 in the WT) can be clearly seen in the figure. The conductivity ratio $\sigma_{K}^{T,WT}/\sigma^{T,MuT}_{K}\sim 12.5$. Conduction via S1-S4 in the WT is almost barrier-less, corresponding to maximum conductivity while, for the mutant, an incoming ion faces an energy barrier of $\sim 4kT$ obstructing entry. Although this barrier is less than those observed in simulations \cite{Tilegenova2019}, conduction in our theory is also inhibited by the loss in conductivity of S4'.

We note that similar results have been obtained experimentally for other KcsA mutants for example,   threonine at S4 has been replaced with alanine, decreasing conductance by a factor of $\sim 17$ at the potential $+100$mV \cite{labro2018inverted}. In addition, the second site, S2, has also been mutated \cite{Tilegenova2019} by substituting glycine with either alanine or cysteine, effectively removing S3, reducing occupancy of S1, and decreasing conduction by a factor of $\sim 32$ at 200mV. We expect the theory to be applicable to these and a wide range of other point mutations.

Our statistical and linear response theory accounts quantitatively for ionic conduction and selectivity in the KcsA and mutant channels used as examples. It takes account of the geometry of individual sites, of long range interactions, and of strongly correlated ionic motion in the SF. In KcsA, conduction is found to occur at almost the rate of free diffusion but with strong selection of K$^+$ over Na$^+$, in accord with experiments. This fast conduction requires the SF to have nearly identical (isoenergetic) binding sites, and optimal values of fixed charge and excess chemical potential at the sites. We find that the Eisenman relations of strong thermodynamic selectivity follow directly from the condition for fast conduction, thereby resolving analytically the long-standing \textit{selectivity-conductivity paradox}. Our results may also bear on the recently-proposed 3-4 ion knock-on conduction mechanism \cite{Koepfer2014,kopec2018direct} because an increased total pore charge would favour 3-4 as the dominant transition. Finally we expect our theory to be of great interest and applicability to artificial nanopores which are designed specifically on transport functionality.

\pagebreak

\onecolumngrid

\section{Physics of selective conduction and point mutations in biological ion channels\\SUPPLEMENTAL MATERIAL}



    We elaborate on the theory developed in the main paper, for those readers who are interested in greater detail. We first discuss the statistical properties including the possible ion configurations in the pore and the link between our free energy and the potential of mean force (PMF). Then we derive the equations governing ionic conductivity at small applied potentials (linear response), and consider three important examples to further illuminate understanding the central figure of the main paper (Figure 3.). Finally we briefly review the important parameters $\Delta\bar\mu_{K,1-4}$ and $\Delta\bar\mu_{Na,1-4}$ and place them in the context of other results.


\section{Configurations in the pore}

The number of possible states corresponding to the total number of possible configurations $\{n_j\}$ of ions binding to the pore can be calculated explicitly. If we consider indistinguishable ions and empty sites (or holes) then we can use the multinomial theorem with each of the $i=1 \cdots S$ species plus empty sites $n_0$ contributing to the total number of states. We note that if the system is reduced to a single species then we can count the number of states using the Binomial theorem. Thus, for $n=\sum_in_i$ ions in a pore with $M$ binding sites, the  number of states can be calculated from, \begin{equation}\mathcal{M}(M,n_1,\cdots n_S,n_0)=\begin{pmatrix}M \\ n_1,\cdots, n_S, n_0 \end{pmatrix} = \frac{M!}{n_0!\prod_i n_i !}\label{Multi},
\end{equation}
where $n_0=M-\sum_i n_i$. It is clear that Eq.\ \eqref{Multi} already accounts for the indistinguishability of ions and is the reason why the $kT\log(n_i)!$ term is included in the expression for the total energy (see (1) of the main text).  Therefore the total number of states $\mathcal{N}$ is given by the sum of this coefficient over all possible numbers of ions and configurations of species inside the pore. Hence, $ \mathcal{N}$ is given by the sum over all possible values of $n_i$ and $n_0$ that satisfy the condition $n_0+\sum_i n_i = M$,

\begin{equation}
    \mathcal{N}=\sum_{n_0+\sum_i n_i = M} \mathcal{M}(M,n_1,\cdots n_S,n_0).
\end{equation}
There are four binding sites in the selectivity filter of KcsA ($M=4$) \cite{Doyle:98a} and so, if we consider a single species of permeant ions, then we find 15 possible states. This number rises to 65 if we consider two conducting  species.

\section{Derivation of the  ionic conductivity}
In this section, we aim to derive the conduction properties of our system in the linear response limit. We consider a small applied electric potential held constant for a long time, and follow Kubo and Zwanzig to investigate the static response using equilibrium statistical mechanics  \cite{kubo1957statistical,zwanzig2001nonequilibrium}.  The equilibrium free energy considered in the main paper (see Eq.\ (5)), is written for the system in electrochemical equilibrium, whence the applied potential contribution is only present if the bulk solutions are asymmetrical. Here, we consider symmetrical solutions and so the free energy with a small applied electrical potential $ G(\{n_j\};n_f;\phi)$ is written as,
\begin{align}
    G(\{n_j\};n_f;\phi)&=\mathcal{E}(\{n_j\})+kT\sum_i\ln n_i!+kT\ln n_0! -\sum_i\sum_m n_{im}[\Delta \bar\mu_{im}+qz_i(\phi^b-\phi^m)+kT \ln(x_i)], 
    \nonumber \\&
    =G(\{n_j\};n_f)+ \sum_i\sum_m n_{im}qz_i(\phi^b-\phi^m).
\end{align}
The ions exit bulk $b$ upon entry to the pore and so $\phi^b$ denotes the electrical potential of the bulk that the ions have exited. We introduce the voltage drop, $\phi$, as defined by $\phi=\phi^L-\phi^R$ and we normalise it such that $\phi^R=0$. The field produced by this voltage drop at each binding site is defined by  $\phi_m=\alpha_{im}\phi$ where $\alpha_m$ is the electrical distance of each binding site for each ion species defined from left to right. Therefore, with this small applied potential the distribution and partition functions become,\begin{align}
    &P(\{ n_j \}; n_f;\phi) = \mathcal{Z}^{-1} \left( \frac{1}{n_0!} \prod_i \frac{(x^b_i)^{n_i}}{n_i!}\right) 
		\times
	\exp\left[\left(\sum_{i=1}^S\sum_{m=1}^M n_{im}\left(\Delta \bar\mu^b_{im}+(\phi^b-\alpha_{im}\phi)z_iq\right)-\mathcal{E}(\{n_j\};n_f)\right)/kT\right]
	\\&
	Z(\{ n_j \}; n_f;\phi) = \sum_{\{n_j\}} \left( \frac{1}{n_0!} \prod_i \frac{(x^b_i)^{n_i}}{n_i!}\right) 
		\times
	\exp\left[\left(\sum_{i=1}^S\sum_{m=1}^M n_{im}\left(\Delta \bar\mu^b_{im}+(\phi^b-\alpha_{im}\phi)z_iq\right)-\mathcal{E}(\{n_j\};n_f)\right)/kT\right].
\end{align}
To simplify notation we shall drop the functional dependence of state space and pore charge  in both the probability and partition functions. The applied potential is directional and hence the voltage drop changes in sign with a positive voltage denoting moving from left to right and vice versa for a negative drop. As a result we introduce parameter $\upsilon^b_{im} =\{1-\alpha_{im},+\alpha_{im}\}$, defined for positive or negative applied potentials.

In the limit of small $\phi$ we can linearise these expressions, 
\begin{align}
    &Z(\phi) \approx Z\left[1\pm \frac{q\phi}{kT}\left\langle \sum_i\sum_m z_in_{im} \upsilon_{im}^b  \right\rangle\right]
\end{align}and,\begin{equation}
    P(\phi) \approx P\left[1\pm\frac{q\phi}{kT} \left(\sum_i\sum_m z_in_{im} \upsilon_{im}^b - \left\langle \sum_i\sum_m z_in_{im} \upsilon_{im}^b \right\rangle\right)\right],
\end{equation}
and they  can be further simplified because the sum over sites can be removed if we introduce an averaged electrical distance $\hat\upsilon^b_i$ for each ionic species, defined by: $\sum_m n_{im}\upsilon^b_{im}=n_i\hat\upsilon^b_i $. Note that this parameter $\hat\upsilon^b_i$ is also averaged over all possible states and so if the sites are evenly distributed, i.e.\ a symmetrical pore, then it is equal to $ 1/2$. Note that the $\pm$ here is present due to the direction and hence sign of $\phi$. If multiple species (possibly with different valence) reside in the pore, then the total influence of $\phi$ can be calculated for each species. In this work we only consider a symmetrical pore, and the effects of varying this parameter will be investigated in future work. 
 
  To linear order the average of any dynamical variable $A$ in an applied field $E$ can be written in terms of the static susceptibility $\chi_{AM}$ \cite{zwanzig2001nonequilibrium}, 
 \begin{equation}
     \langle A \rangle_E = \langle A \rangle + \chi_{AM}E.
 \end{equation}
The response or $\chi_{AM}$ describes the average linear response produced by the applied field, and contains information about the coupling $M$ to the field. 

We shall calculate the static susceptibility for the whole pore and the binding site $m$, in response to the field generated by the scalar potential $\phi$. As a result, in accordance with Kubo, we shall be looking for a relation of the form,
\begin{equation}
         \langle n \rangle_\phi = \langle n \rangle + q\chi\phi.
\end{equation}
where $\chi$ has units of [$kT^{-1}$] or [$(m^3kT)^{-1}$] depending on if it is the response for particle number or density. 

If we first consider the the full pore, then we must average over total particle number inside the pore $\left\langle  n_{i}  \right\rangle_{\phi}$,
 \begin{equation}
         \left\langle  n_{i}  \right\rangle_{\phi} = \sum_{\{n_j\}}  n_{i} P(\phi)
        = \left\langle n_{i}  \right\rangle \pm \frac{q\phi}{kT}\left(\left\langle n_{i}\left(\sum_i z_in_{i}\hat\upsilon_i^b\right)\right\rangle-\left\langle \sum_i z_in_{i}\hat\upsilon_i^b\right\rangle \left\langle n_{i}\right\rangle\right),
 \end{equation}
whereas at each site  we must average over total particle number inside the site $\left\langle  n_{im}  \right\rangle_{\phi}$
\begin{equation}
    \left\langle  n_{im}  \right\rangle_{\phi} = \sum_{\{n_j\}}  n_{im} P(\phi)
        = \left\langle n_{im}  \right\rangle \pm \frac{q\phi}{kT}\left(\left\langle n_{im}\left(\sum_i z_in_{i}\hat\upsilon_i^b\right)\right\rangle-\left\langle \sum_i z_in_{i}\hat\upsilon_i^b\right\rangle \left\langle n_{im}\right\rangle\right).
\end{equation}
The resulting static density susceptibilities are written as, 
\begin{align}
&          \chi_{im} = \left(\left\langle n_{im}\left(\sum_i\sum_m z_in_{im}\upsilon_{im}^b\right)\right\rangle-\left\langle \sum_i\sum_m z_in_{im}\upsilon_{im}^b\right\rangle \left\langle n_{im}\right\rangle\right)\frac{1}{V_{im}kT}. 
\\&
\chi_{i} = \left(\left\langle n_{i}\left(\sum_i\sum_m z_in_{im}\upsilon_{im}^b\right)\right\rangle-\left\langle \sum_i\sum_m z_in_{im}\upsilon_{im}^b\right\rangle \left\langle n_{i}\right\rangle\right)\frac{1}{VkT} 
\end{align}
We note that $V_{im}$ is the site volume for each respective species $i$. The total volume of the pore is constant, but we allow the volume of each site to vary with species such that we can write $\sum_m V_{im} \equiv V$. In both of these susceptibilities we have explicitly included correlations between ions of different species and between ions at different sites because both of these may be non-negligible. We can also write these functions using equilibrium  statistical mechanics,
\begin{align}
    &        \chi_{im} = \left(\hat\upsilon_{i}^bz_i  \frac{\partial \langle n_{i}\rangle }{\partial \Delta \bar\mu_{im}}+\sum_{j\neq i}^S\hat\upsilon_{j}^bz_j\frac{\partial \langle n_{j}\rangle}{\partial \Delta \bar\mu_{im}} \right)\frac{1}{V_{im}kT} 
    \label{Eqn:SiteSusc}    \\&
    \chi_{i} = \left(\hat\upsilon_{i}^bz_i \sum_{m=1}^M\frac{1}{V_{im}} \frac{\partial \langle n_{i}\rangle }{\partial \Delta \bar\mu_{im}}+\sum_{j\neq i}^S \hat\upsilon_{j}^bz_j \sum_{m=1}^M \frac{1}{V_{im}} \frac{\partial \langle n_{j}\rangle}{\partial \Delta \bar\mu_{im}} \right)\frac{1}{kT}  \label{Eqn:TotalSusc}.
\end{align}
Clearly, the total species susceptibility is written as $\chi_i=\sum_m \chi_{im}$, which can be summed over species. 

In total, the pore can be represented by a series of connected binding sites, and charge must be conserved meaning that current can only flow through the pore if all sites are conducting. Thus the pore is analogous to resistors in series and the resistivity $\mathcal{R}_{im}$ and conductivity $\sigma_{im}$ can be determined at each binding site. The reciprocal values of $\sigma_{im}$ can be summed to obtain the total species conductivity. According to linear response theory the conductivity of a system can be calculated from the Einstein relation which relates the diffusivity $D$ to the susceptibility. This is derived in \cite{kubo1957statistical,kubo1966fluctuation} for a system of volume $V$ as $\sigma = q^2D\langle(\Delta n)^2 \rangle/(VkT)$ where the static density susceptibility is defined as $\chi=\langle(\Delta n)^2 \rangle/(VkT) $. However, in our system $\chi_{i}$ or total variance in particle density may remain non-zero and even large when a binding site becomes non-conductive, which is in clear violation of Kirchoff's laws. Clearly, then, we need to determine the conductivity at each binding site in order to compute the total. Hence, we use the Einstein relation to calculate site-conductivity $\sigma_{im}$ in terms of $\chi_{im}$  Eqn. \eqref{Eqn:SiteSusc},\begin{equation}\label{Eqn:Cond}
\sigma_{im}=z_iq^2\chi_{im}D_{im}.
\end{equation}   Total species conductivity $\sigma_i^T$ can be computed, and this must be summed over species in order to obtain the total conductivity of the pore, \begin{equation}
    \frac{1}{\sigma_i^T}=\sum_m \frac{1}{\sigma_{im}}, \quad 
    \sigma_T=\sum_i \sigma_i^T.
\end{equation}
We note that if all ion types and sites become indistinguishable, then each site conductivity and hence $\sigma^T$ becomes proportional to the variance in total particle number as derived in \cite{kubo1957statistical,kubo1966fluctuation}. 

This result is in full agreement with numerous experimental examples within the literature. For example, point mutations affecting conduction in K$^+$ or NaK channels \cite{derebe2011tuning}, or C-type inactivation in KcsA  \cite{Xu2019}. It is clear from Eqn. \eqref{Eqn:SiteSusc} that the conductivity is being written for a strongly correlated system, including  single particle (self) and inter-particle correlations (perhaps between ions of different species). Furthermore, local pore volume can also determine conductivity/selectivity and the effects of mutation can be described. 

Conductivity and hence electrical current can be defined at each site, \begin{equation}
    \mathcal{G}_{im}= \frac{A_{im}}{L_{im}} \sigma_{im}, \quad     I_{im}=\mathcal{G}_{im}\mathcal{V}_m
\end{equation}
where $A_{im}$, $L_{im}$ and  $\mathcal{V}_m$  are surface area, length and applied voltage at each site $m$ respectively. Since total current across the whole pore equals current across each site, we can  write the total current as,
\begin{equation}
    I_i=\mathcal{G}_i^T \mathcal{V}, \text{ where: } \mathcal{G}_i^T= \left(\sum_{m}^M\frac{1}{\mathcal{G}_{im}}\right)^{-1}. 
\end{equation}
Here $\mathcal{V}=\sum_m \mathcal{V}_m$ is the total applied voltage. 

If we consider a pore consisting of $M$ sites of identical geometry, and constant diffusivity throughout the pore, then the current simplifies to,
\begin{equation}
    I_i =  \frac{z_iq^2}{L_{im}^2} D_i^c \left(\sum_{m=1}^M\frac{1}{\tilde\chi_{im}}\right)^{-1}  \mathcal{V}, 
\end{equation} 
where $L_{im} = \left< L_{im}\right>$ is the average length of the binding site and $\tilde\chi_{im}$  is the number susceptibility equal to $\chi_{im}$ multiplied by the volume at each site (which is equal for each species and site in this instance). Furthermore, if each site shares binding affinity such that, $\tilde\chi_{i}\equiv\tilde\chi_1=\tilde\chi_M$, then one can write, 
\begin{equation}
        I_i =  z_iq^2\frac{D_i^c}{L_{im}^2} \tilde\chi_{i}  \mathcal{V}.
\end{equation}

Let us consider our example of KcsA with $4$ sites and a single conducting species. If site 4 has volume $V_4$ and sites 1-3 share volume $V_{m}$, then by introducing parameter  $\beta$ defined as: $V_4=\beta V_{m}$,  we can write,
\begin{equation}
       \sigma^T=\frac{Dzq^2}{V_m}\frac{\tilde\chi_1\tilde\chi_2\tilde\chi_3\tilde\chi_4}{\beta \tilde\chi_1\tilde\chi_2\tilde\chi_3+ (\tilde\chi_1\tilde\chi_2\tilde\chi_4+\tilde\chi_1\tilde\chi_3\tilde\chi_4+\tilde\chi_2\tilde\chi_3\tilde\chi_4)}.
\end{equation} Interestingly when we consider the total conductance then this change in volume only affects the conductance if the length is varied. Thus, if we introduce $L_4=\gamma L_{m}$ we can write total conductivity as,
\begin{equation}\label{Eqn:ConductanceFit}
    \mathcal{G}^T=\frac{Dzq^2}{L_{m}^2}\frac{\tilde\chi_1\tilde\chi_2\tilde\chi_3\tilde\chi_4}{\gamma \tilde\chi_1\tilde\chi_2\tilde\chi_3+ (\tilde\chi_1\tilde\chi_2\tilde\chi_4+\tilde\chi_1\tilde\chi_3\tilde\chi_4+\tilde\chi_2\tilde\chi_3\tilde\chi_4)}.
\end{equation}

In the main paper we fit experimental current-voltage relations found for wild-type (WT) KcsA and a mutant (MuT). Mackinnon et. al. \cite{zhou:2004} states that the cysteine replaces threonine without significantly altering the side-chain volume. As a result we assume that all four sites S1-4 in the WT and S1-3 in the MuT have the same volumes, and consider S4 in the MuT to be slightly perturbed with $\gamma=1.2$.  Furthermore we assume that both channels remain fully symmetrical with respect to the applied voltage, meaning that $\hat\upsilon_{m}^b=1/2$.

\section{Conductivity Examples}
The occupancy and conductivity highlighted by Fig.\ 3 of the main paper, exhibit rich and complex behaviour. To explore and explain these phenomena we now consider 3 important examples outlining the conductivity-occupancy relationship, and the effects of allowing more ions and ion types inside the pore. Note that we assume KcsA geometry with 4 sites and a total channel length and radius of 12\AA\; and 2\AA\; respectively. Although not necessary, we also consider the limit where all site volumes are identical and equal to the total pore volume divided by the number of sites, $V_{im}\equiv V^c/M$, and where the diffusivities are also equal in each site.

\begin{figure}[t]
    \centering
    \includegraphics[width=0.75\linewidth]{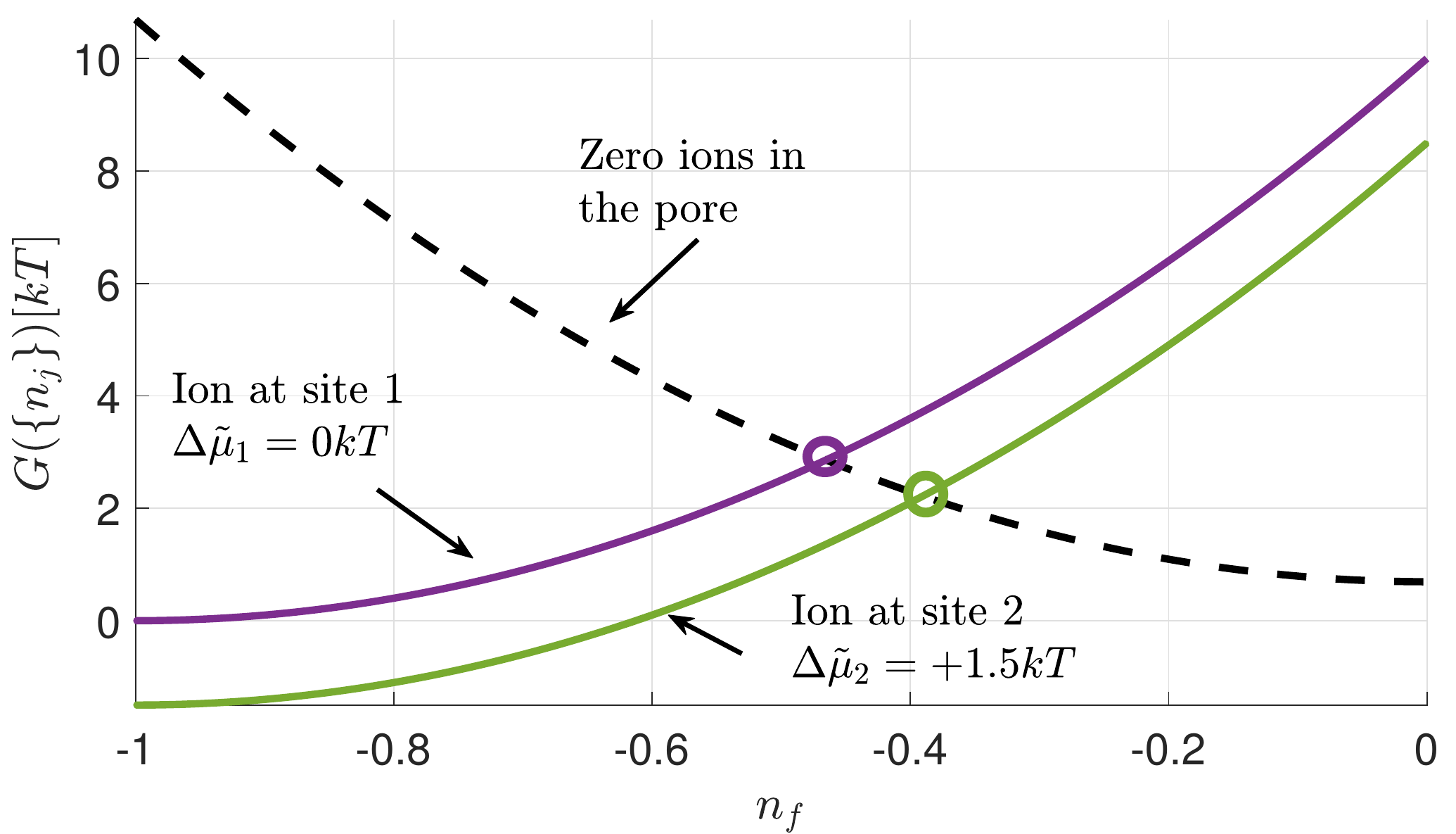}
    \caption{Free energy $G$ \textit{vs.} the fixed charge $n_f$.  The black dashed curve represents the state without ions; the purple and green curves represent states with an ion at site 1 or 2 respectively. The purple and green circles highlight the degeneracy knock-on condition between states with 1 and 0 ions (see Eqn. \eqref{eq:Ex1Condition}).   }
    \label{fig:Ex1a}
\end{figure}

\subsection{Example 1: One species, one ion, multiple binding sites}
In this example we investigate the system under the 0-1 ion transport mechanism, for  a pore with two binding sites. We consider the binding affinity in site 1 to be fixed but mutate site 2 and thus allow $\Delta\tilde\mu_{2}=\Delta\bar\mu_2+kT\log(x)$ to vary in accordance with Fig.\ 3 of the main paper. The aim  is to describe the effect on the conduction mechanism of the interplay of distinguishable binding sites. In this system the partition function comprises of one empty and two singly-occupied states. We denote the number of ions and their location within the free energy,
\begin{align}
    Z=e^{-G(0)/kT}+e^{-G(1K,m=1)/kT}+e^{-G(1K,m=2)/kT}
    \end{align}

In  Fig. \ref{fig:Ex1a} we consider the energy spectra $vs.$ charge of the pore $n_f$. The black dashed curve represents the state without ions; the purple and green curves represent states with an ion at site 1 and 2 respectively. When the ion is in site 2 we showcase the spectra when $\Delta\tilde\mu_2=1.5kT$. Coloured circles highlight the occurrence of degeneracy between the zero and strongly occupied states. At each of these points we find that the respective energy barrier is zero i.e. $\Delta G =0$. Here, the system is degenerate because either state is equally probable and, under certain conditions that we will elucidate later, may yield conduction. Due to the difference in $\Delta\tilde\mu$, splitting between the single ion states occurs, resulting in a shift of the degeneracy point. However, it is clear that when such a degeneracy occurs at a higher energy than another i.e. the purple curve in this figure; then the probabilities of the states in question  are smaller than the probability for the ion to be in site 2. Thus we can establish the following energy barrier at each degeneracy point,\begin{equation}\label{eq:Ex1Condition}
  \Delta G_m = \Delta\mathcal{E}+ kT\log[(n+1)/n_0] -\Delta\bar\mu_m -kT\log(x)=0.
\end{equation}

\begin{figure}[t]
    \centering
    \includegraphics[width=0.75\linewidth]{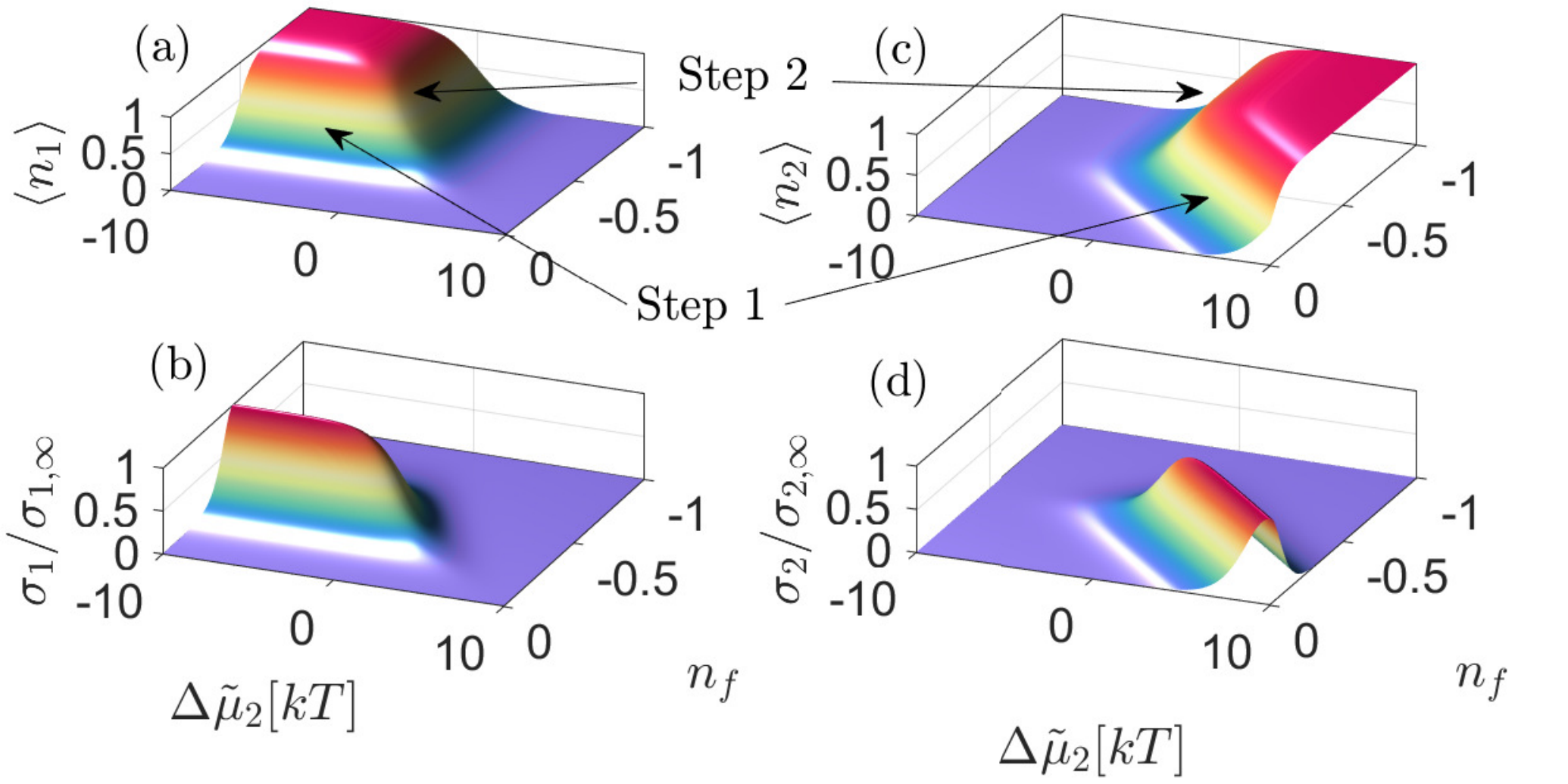}
    \caption{Occupancy (a,c) and normalised conductivity (b,d) into site 1 and 2 respectively, are plotted $vs.$ $n_f$ and $\Delta\bar\mu_2$. Two steps in occupancy form, one when an ion is entering the pore i.e.\ facing the $\Delta\bar\mu_2$ axis and the second when an ion is in the pore but is either in site 1 or 2. The former step corresponds to the peak in conductivity into the site.  }
    \label{fig:Ex1b}
\end{figure}

Let us consider the occupancy and conductivity of each site (see Fig. \ref{fig:Ex1b}). Occupancy ((a) and (c)) forms a step-like function $vs.$ both  $\Delta\tilde\mu_2$ and  $n_f$ labelled in the figure  as step 1 and 2 respectively. Step 1 also corresponds to a step in total particle number, but step 2 does not because here the single ion states are always favoured and it is just the pore favouring site 1 or 2. The conductivity forms a sharp peak close to the midpoint of step 1. Therefore, we can conclude that conduction maximises at the degeneracy condition \eqref{eq:Ex1Condition} and when the site is favoured. When a  difference between site affinities exists, conduction may still occur but it is reduced.  

\begin{figure}[t]
    \centering
    \includegraphics[width=0.75\linewidth]{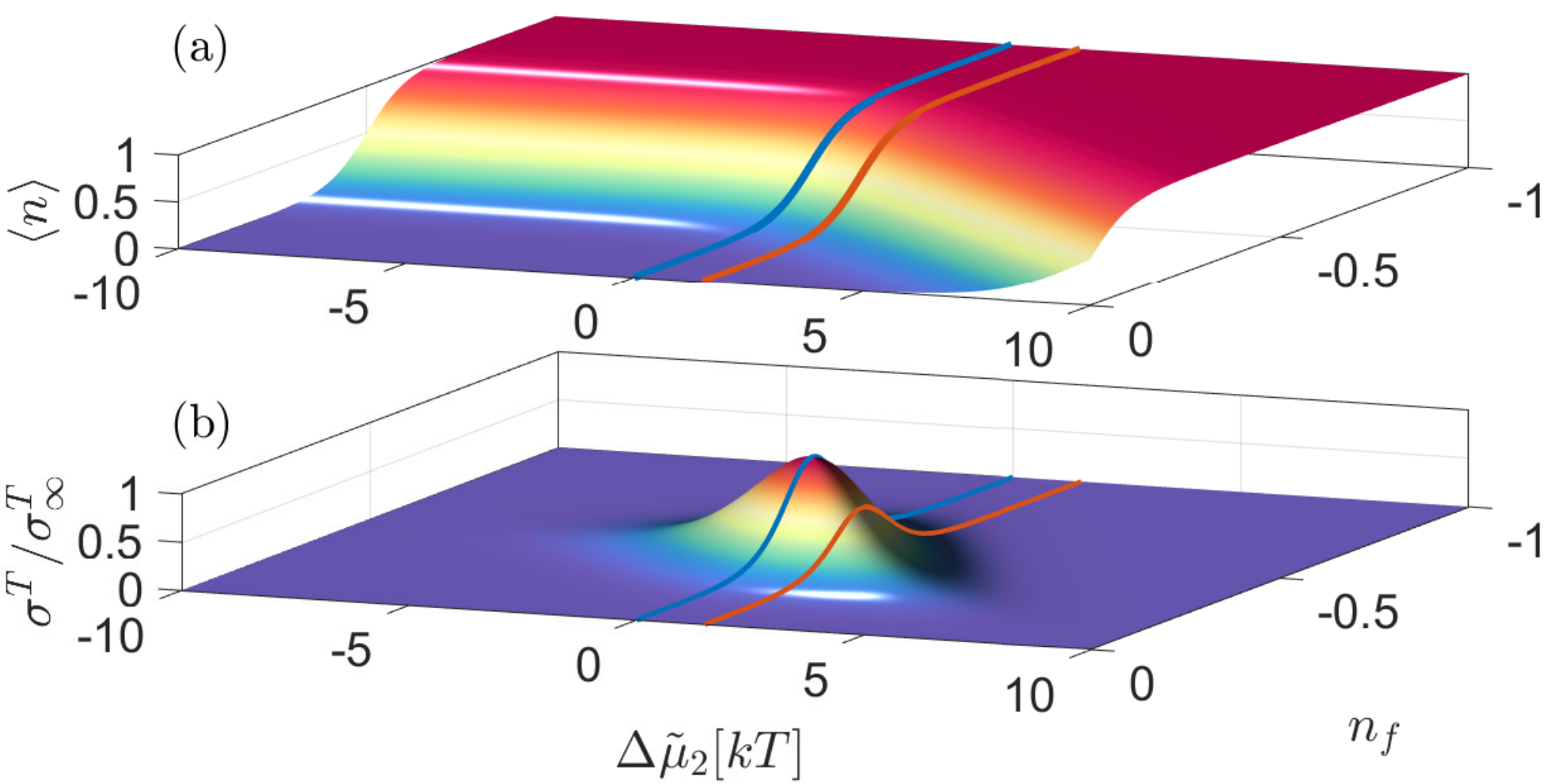}
      \includegraphics[width=0.75\linewidth]{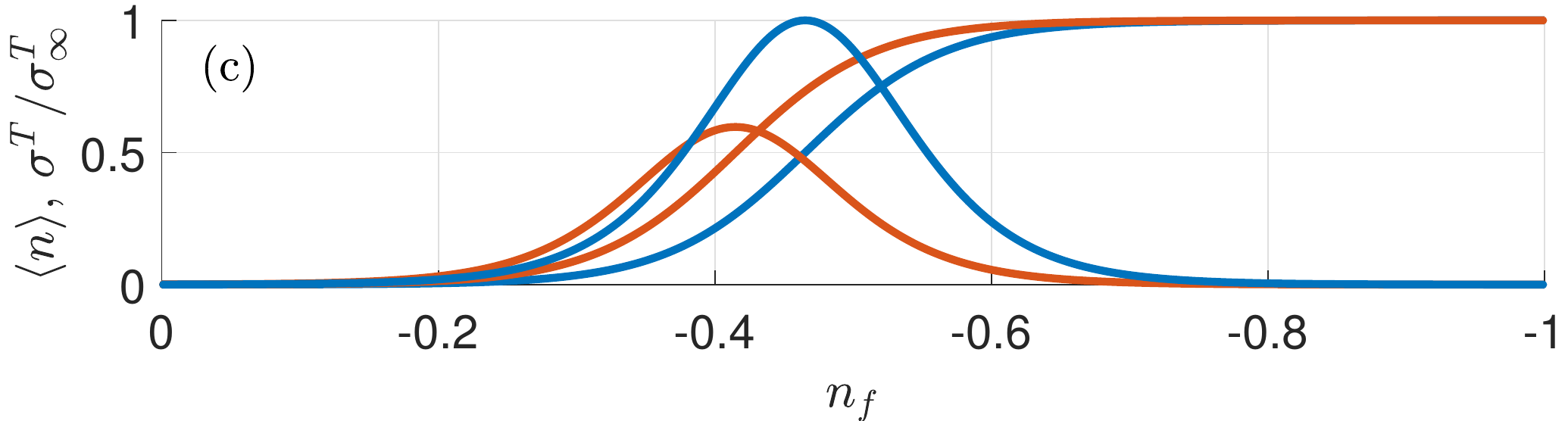}
    \caption{Occupancy (a) and normalised conductivity (b) plotted $vs.$ $n_f$ and $\Delta\tilde\mu_2$. Occupancy forms a step that varies dependent on $n_f$ and $\Delta\tilde\mu_2$ only when site 2 is more favoured. The midpoint of this step corresponds to the degeneracy between states given by Eqn. \eqref{eq:Ex1Condition}. Conductivity forms a peak around this point if the two binding sites also share affinity i.e. $|\Delta\tilde\mu_1-\Delta\tilde\mu_2| \sim 0$. In analogy with quantum transport through quantum dots the phenomena along the $n_f$ axis can be attributed to ionic Coulomb blockade. }
    \label{fig:Ex1c}
\end{figure}

The total conductivity and occupancy are considered in Fig.  \ref{fig:Ex1c}. Panels (a) and (b) show the total occupancy and normalised conductivity through the pore $vs. n_f$ and $\Delta\tilde\mu_2$. It is clear from both figures that the $\Delta\bar\mu_2=0$ plane is important because it is around this that conductivity is peaked and where occupancy starts to vary due to the growing importance of the state with the ion occupying site 2.  

Total conductivity is given by the inverse of the sum of the reciprocal conductivities at individual  sites. For this example we can write,
\begin{equation}\label{eq:Ex1sigma}
    \sigma^T=\frac{\sigma_1\sigma_2}{\sigma_1+\sigma_2}
\end{equation}
where we remind the reader that each conductivity is proportional to the susceptibility into the given site.  Therefore as the difference in affinity between sites grows i.e. the limit $|\Delta\tilde\mu_1-\Delta\tilde\mu_2|\gg 0$ the susceptibility and conductivity into the least favoured site tends to zero and the total conductivity drops to zero. Thus, in contrast to conduction into the individual sites, conduction now occurs close to the degeneracy condition \eqref{eq:Ex1Condition} provided that the difference in site affinities is small $|\Delta\tilde\mu_1-\Delta\tilde\mu_2| \sim 0$. In fact, it is not exactly zero because we have omitted a small factor $\epsilon$ due to the entropy of mixing of binding sites. 

To gain further insight, we now consider the conductivity and occupancy of site 1 only and examine the limit that site 1 is more favoured than site 2 i.e. $\Delta\tilde\mu_1\gg\Delta\tilde\mu_2$,
\begin{align}
   & \langle n_1 \rangle = \frac{e^{-\Delta G_1/kT}}{1+e^{-\Delta G_1/kT}}
    \\&
\sigma_1 \propto  \frac{e^{-\Delta G_1/kT}}{(1+e^{-\Delta G_1/kT})^2}.
\end{align} 
Here, $\Delta G_m=G(1K,m)-G(0)$. Therefore it is evident that conductivity into each site maximises at the midpoint of the occupancy step, i.e.\ $\Delta G_m=0$ when this is the favoured site. If we consider the total occupancy and conductivity then we would find, 
\begin{align}
   & \langle n \rangle \rightarrow  \langle n_1 \rangle
    \\&
\sigma \rightarrow 0.
\end{align}
If we consider the total conductivity and occupancy when the site affinities are equal i.e. $\Delta G_1=\Delta G_2= \Delta G$ then we can write, 
 \begin{align}
   & \langle n \rangle = \frac{2e^{-\Delta G/kT}}{1+2e^{-\Delta G/kT}}  
    \\&
\sigma \propto \frac{e^{-\Delta G/kT}}{(1+2e^{-\Delta G/kT})^2}.
\end{align}Therefore, it is clear the the peak of the total conductivity is slightly offset from the condition $\Delta G=0$, due to a factor of $kT\ln(2)$. This occurs due to the degeneracy between sites which results in multiple (2 in this instance) possible transitions.  However, these contributions are often small and so we neglect them and write the following two conditions for maximal conductivity,
\begin{align}
&\Delta G= 0 \label{Condition1}
\\&
\Delta\bar\mu_{i1}\sim \Delta\bar\mu_{im'}, \in m'=2,\cdots, M \label{Condition2}
\end{align}

Finally it is worth noting that here the fixed charge here plays an identical role to that of the gate voltage in quantum dots and transistors. As a result the conductivity peak and occupancy step correspond to ionic Coulomb blockade.  This ionic analogue of electronic Coulomb blockade has been already been identified as an important classical ionic transport process found in artificial and biological channels \cite{Krems:13,Kaufman:15,Feng:16,kavokine2019ionic}.  This is clearly observed in (c) of Fig. \ref{fig:Ex1c} where two traces at $ \Delta\tilde\mu_2=0$ and $\Delta\tilde\mu_2=1.5kT$ produced shifted conductivity peaks and Coulomb steps in occupancy. 

\subsection{Example 2: One species, two ions, two binding sites}

\begin{figure}[t]
    \centering
    \includegraphics[width=0.75\linewidth]{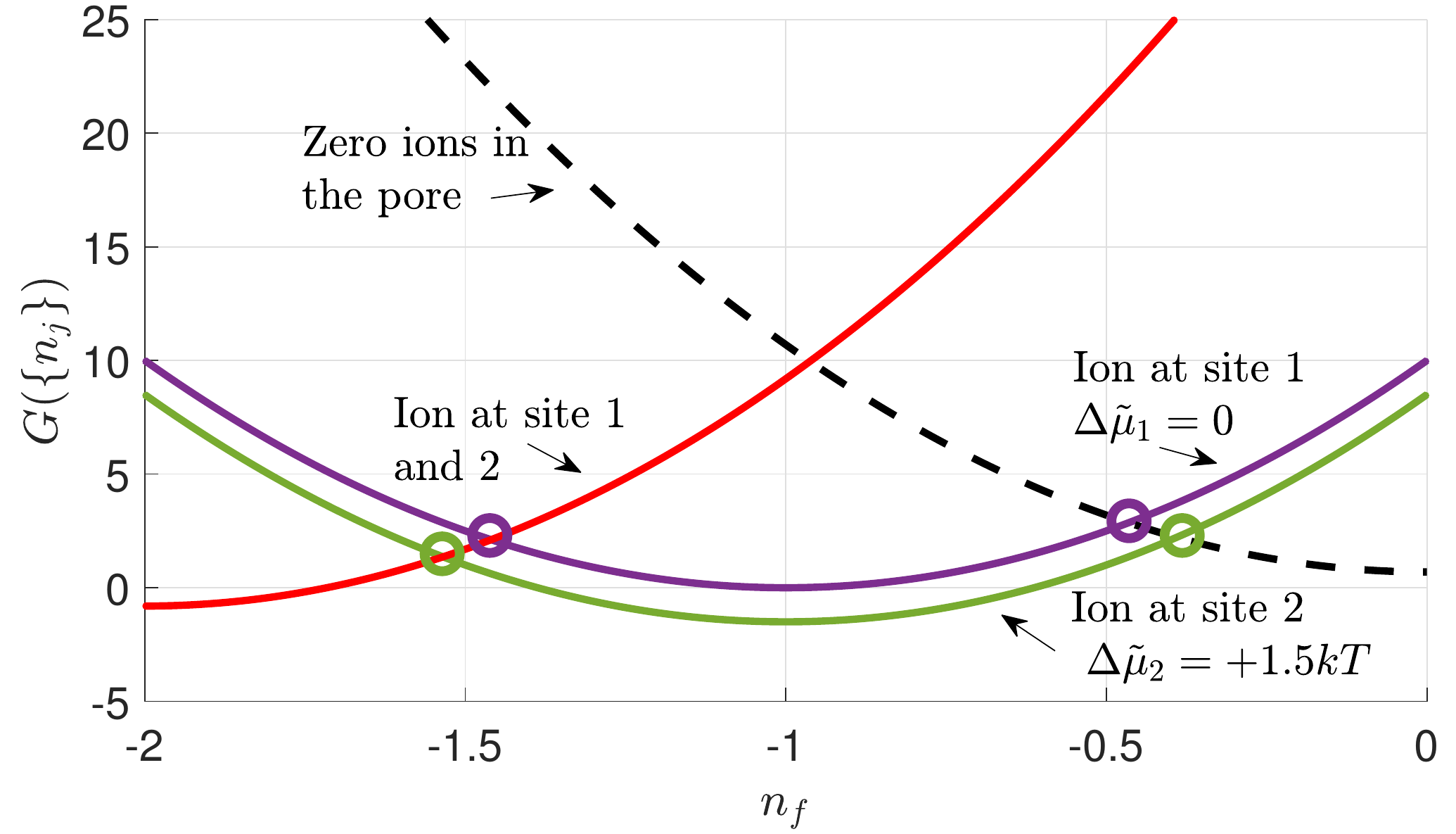}
    \caption{(a) Energy spectra $vs.$ $n_f$. The purple and green curves represent the single-ion states with occupation at sites 1 or 2, respectively; the red and black curves represent the dual-ion and zero-ion states. Coloured circles highlight the degeneracy condition.  }
    \label{fig:Ex2a}
\end{figure}

In this next example we shall restrict ourselves to a single species, but allow up to two ions inside the pore. The aim is to investigate changes in conductivity when varying the number of ions inside the pore and the conduction mechanism  i.e. 0-1 to 1-2 transitions through the pore.  The parameters are identical to those used in the previous example. The partition function is slightly extended to include a fourth double ion state, 
\begin{align}
    Z=e^{-G(0)/kT}+e^{-G(1K,m=1)/kT}+e^{-G(1K,m=2)/kT}+e^{-G(2K,m=1,2)/kT}
    \end{align}
In Fig. \ref{fig:Ex2a} we plot the energy spectra again $vs.\, n_f$. Purple and green curves correspond respectively to single ions at sites 1 and 2 when $\Delta\tilde\mu_2=+1.5kT$. The black dashed line is again the state with zero ions and the red spectrum corresponds to ions in both sites and again $\Delta\tilde\mu_2=+1.5kT$. We covered the first transition in the earlier example and so here we shall focus on the second. 

\begin{figure}[t]
    \centering
    \includegraphics[width=0.75\linewidth]{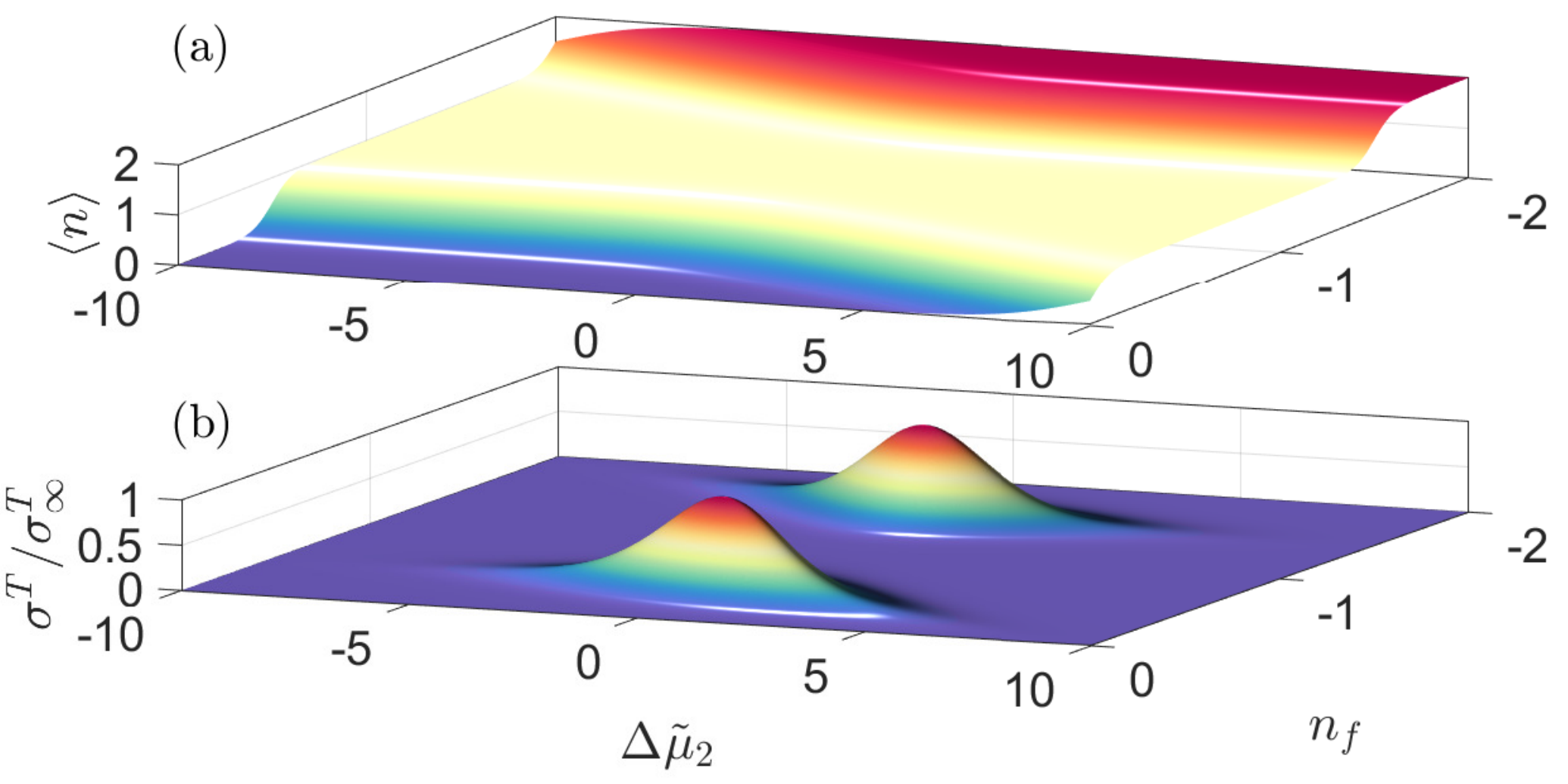}
        \includegraphics[width=0.75\linewidth]{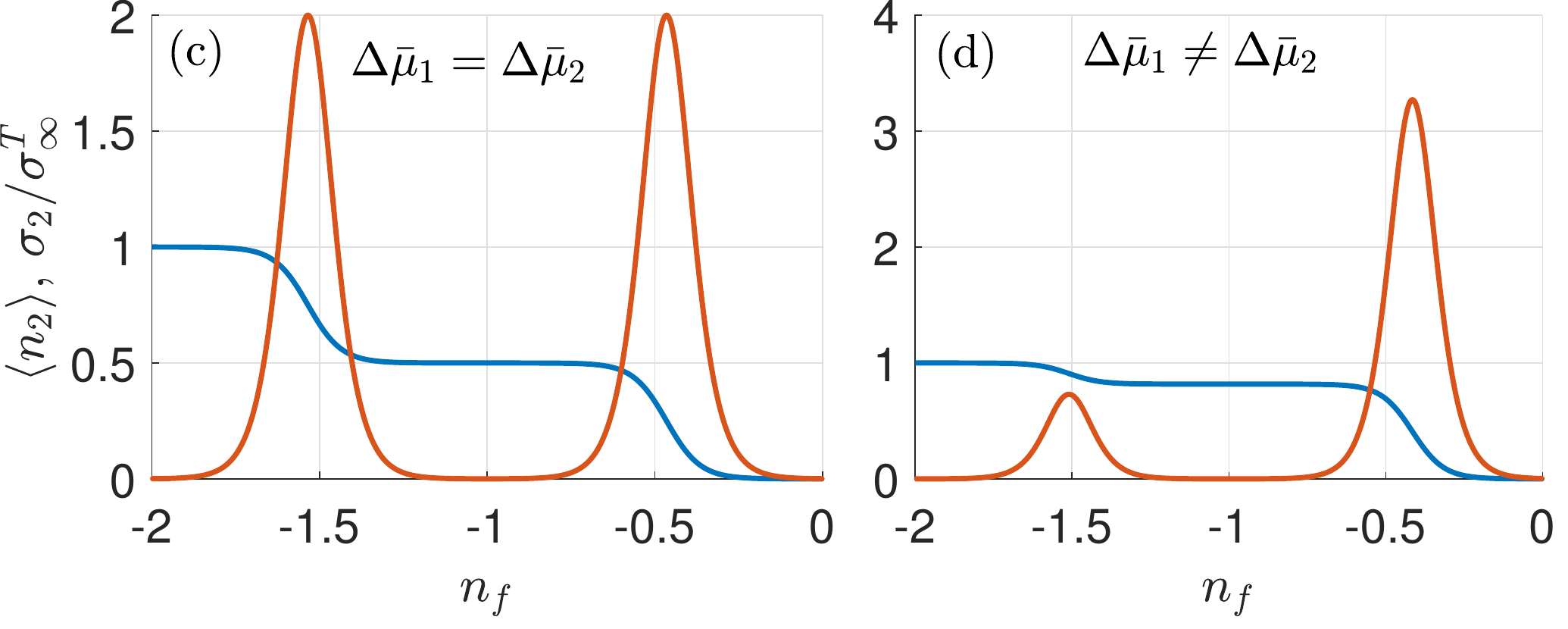}
    \caption{Occupancy (a), normalised conductivity (b) traces when the sites share affinity (c) and there is a favoured site (d). Each conductivity peak again forms close to the degeneracy condition or mid-point of occupancy step, and the frequency between peaks can be calculated from the difference between energy levels. Both functions share an anti-symmetry along the $\Delta\bar\mu_2$ plane because, if either site is full, a transition is only possible to the other one. When sites share affinity the total conductivity involves transitions at both sites and so each site shares conductivity resulting in a smaller total. If an energy difference exists between sites then conductivity into each site is only possible provided that the site is empty. As a result in (d) conductivity is significantly reduced in the 1-2 transition because it is almost filled. If it was filled then the conductivity would be negligible. }
    \label{fig:Ex2b}
\end{figure}

We consider the occupancy and fluctuations in (a) and (b) of Fig. \ref{fig:Ex2b} respectively. Both functions have an anti-symmetrical dependence on the $\Delta\tilde\mu_2=0$. This is because after the first transition, and unless $\Delta\tilde\mu_1\sim \Delta\tilde\mu_2$, one site is more likely to be occupied than the other. To observe this effect we have plotted the occupancy and conductivity of site 2 when the sites  share affinity (c) and when site 1 is more favoured by 1.5kT (d).  When sites share binding affinity the movement of ions into site 2 is equally probable for both the 0-1 and 1-2 transitions. As a result the conductivity peaks and occupancy steps are equal in magnitude. However, when site 2 is favoured in (d) we can clearly see the effects of the energy barrier prohibiting the ion from entering site 2 if the pore already contains an ion because its conductivity is reduced. 

We can now also investigate the periodicity of the peaks along the $n_f$ plane. Since $\Delta\tilde\mu$ is unchanged with differing states of the system the periodicity is largely defined by the difference in the longer-range electrostatic interactions of the pore  $\Delta\mathcal{E}$.   Thus with the approximation used corresponds to the renormalised energy difference in the absence   of  charging  effects:  $2U_c$ \cite{Beenakker:91}. This is easily seen in Fig \ref{fig:Ex2a} if you consider the energy difference between the red and green curves during the first transition.

\subsection{Example 3: Two species, multiple ions, multiple binding sites}
\begin{figure}[ht]
    \centering
    \includegraphics[width=0.75\linewidth]{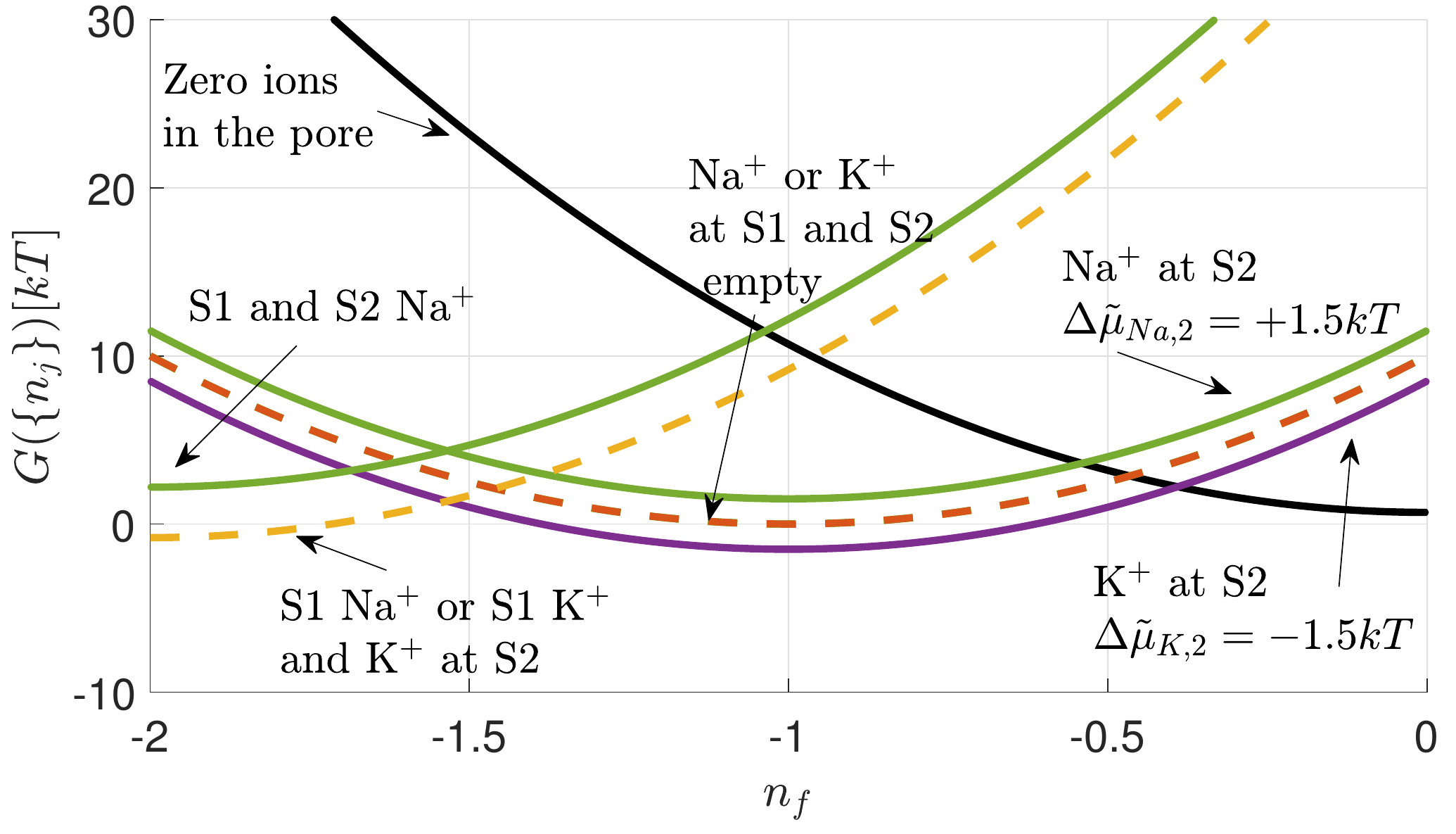}
    \caption{Energy Spectra $vs.$ $n_f$. The solid purple and green curves correspond  to pure K$^+$ and Na$^+$ states, the dashed orange represents either K$^+$ or Na$^+$ at site 1, the dashed yellow line is  either K$^+$ or Na$^+$ at site 1 and K$^+$ at site 2, and finally the black line denotes the zero-ion state. }
    \label{fig:Ex3a}
\end{figure}

In our third example we consider the effects of competition inside the pore by allowing for two conducting species, sodium Na$^+$ and potassium K$^+$.  We take: $U_c=10kT$, $x_K=x_{Na}$,  $ \Delta\tilde\mu_{1K/Na}=0$ for both K$^+$ and Na$^+$ and allow $\Delta\Delta\tilde\mu_{K,Na;S2}$ to vary (thus mimicking the conditions used in Fig.\ 3 of the main paper). In this system the partition function is comprised of the following states,
    \begin{align}
    Z&=e^{-G(0)/kT}+e^{-G(1K,m=1)/kT}+e^{-G(1K,m=2)/kT}+e^{-G(1Na,m=1)/kT}+e^{-G(1Na,m=2)/kT}+
    \nonumber \\ &
    e^{-G(2K,m=1,2)/kT}+e^{-G(2Na,m=1,2)/kT}+e^{-G(1K,m=1;1Na,m=2)/kT}+e^{-G(1K,m=2;1Na,m=1)/kT}
    \end{align}    
We start by considering the energy spectra in Fig.\ \ref{fig:Ex3a}.    The solid purple and green curves correspond  to pure K$^+$ and Na$^+$ states, the dashed orange represents either K$^+$ or Na$^+$ at site 1, the dashed yellow line is  either K$^+$ or Na$^+$ at site 1 and K$^+$ at site 2, and finally the black line denotes the zero-ion state.  There is a major difference to the earlier examples because the energy spectra can now undergo splitting due to the ion type in addition to the configuration i.e. sites occupied. As outlined in the central result of the main paper, selectivity at each site can be determined for a fixed $n_f$ and in the case of symmetrical solutions it yields the Eisenman relation    $|\Delta\tilde\mu_{im}-\Delta\tilde\mu_{jm}|$. We also classify higher order $(n>1)$ states into either pure i.e. 2K$^+$ or mixed 1K$^+$1Na$^+$ states. This is apparent from the dashed purple curve which is the lowest-energy mixed state (the other is not shown) corresponding to K$^+$ at site 2 and Na$^+$ at site 1. As a result mixed species conduction can now occur under specific parameters via pure or mixed states. 

\begin{figure}[ht]
    \centering
    \includegraphics[width=0.75\linewidth]{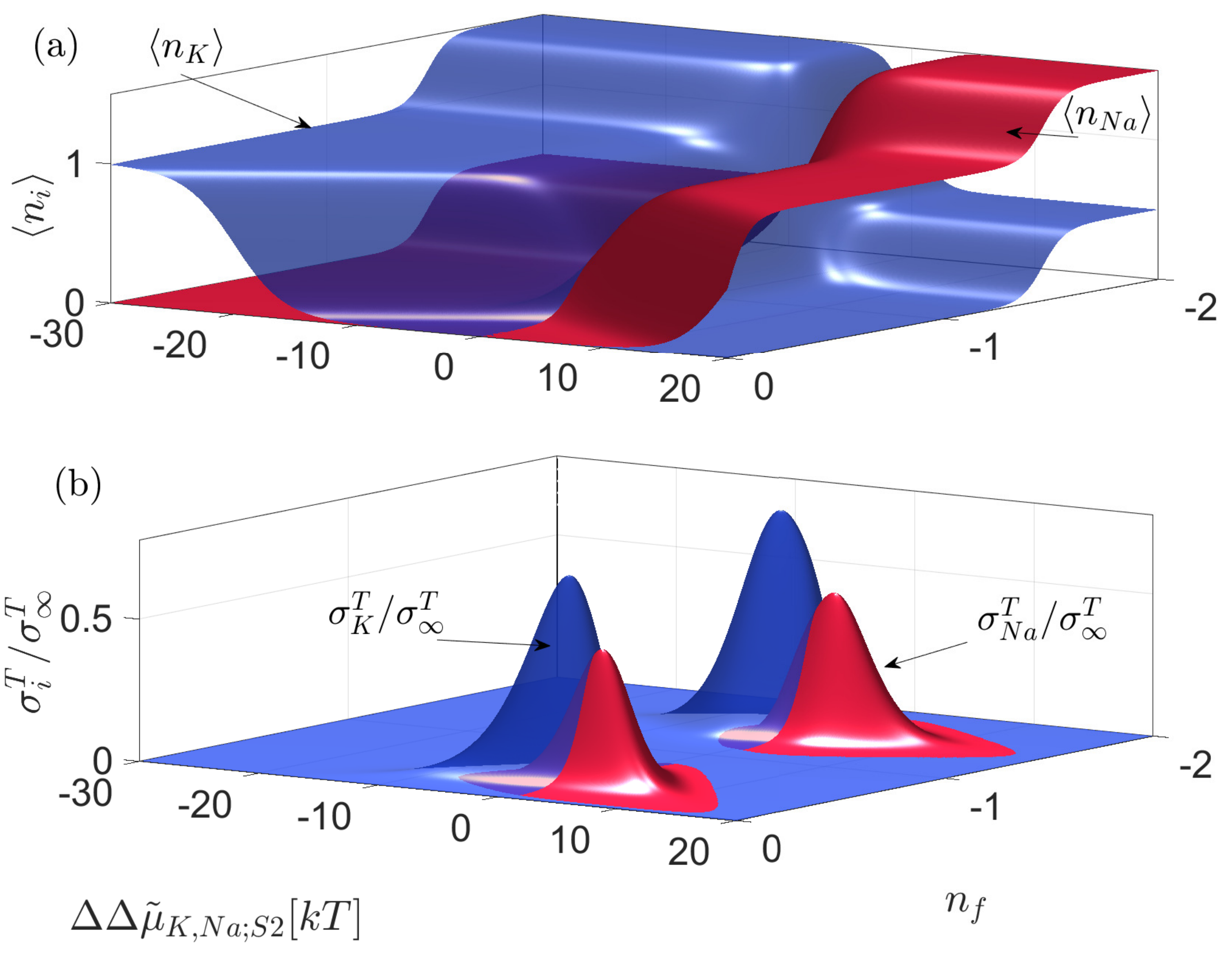}
    \caption{K$^+$ (blue) and Na$^+$ (red) occupancy (a), and normalised conductivity (b)  plotted $vs.$ $n_f$ and $\Delta\Delta\bar\mu_{K,Na;2}$.   }
    \label{fig:Ex3b}
\end{figure}

In Fig. \ref{fig:Ex3b} we consider the total occupancy (a) and conductivity (b) through the pore. As before, an ion can enter any free site. Thus conduction is again governed by which site has the higher affinity. However, this is not the complete picture because sites can be selective and by differing amounts.  Thus in the limit that an ion can only enter site $m$, and this site is strongly favoured for species $i$, one would expect conduction to occur at the degeneracy condition for this site and species (again neglecting the small mixing term). So in our example K$^+$ maximally conducts when $\Delta\tilde\mu_2 \sim 1.5kT$. If the sites are not strongly selective or if selectivity differs at some sites meaning that multiple possible conduction events are possible then conduction is not maximal unless the system is completely isoenergetic and non-selective.

So we can conclude that conduction is governed by strict conditions. If we are interested in maximal and selective conduction for ions of species $i$ then we must require that the following conditions are satisfied,\begin{align}
 &\Delta G_m =0. 
 \\&
 \Delta\bar\mu_{i1}\sim \Delta\bar\mu_{im'}, \in m'=2,\cdots, M 
 \\&
  \Delta\bar\mu_{im} \gg \Delta\bar\mu_{jm}, \in m=2,\cdots, M. \label{Condition3} 
\end{align}

 \section{Relation to PMF}
 
 To derive a relationship between our discrete space free energy $G$ and the continuous multi-ion PMFs calculated from simulation we shall consider the statistical theory developed by Roux \cite{Roux1999,Im:00}. In this work the binding factor $\mathcal{B}_n$  was introduced, representing the ratio of probabilities to have $n$ and $0$ ions within the pore. We shall limit ourselves to a single species occupying up to $M$ binding sites in the pore. In Roux's theory the binding factor was explicitly formulated in terms of the multi-ion PMF $\mathcal{W}$,
 \begin{equation}
     \mathcal{B}_n=\frac{(\bar\rho^I)^n}{n!}\times \int_{c}d\mathbf{r}_1 \cdots \int_c d\mathbf{r}_n e^{[n\bar\mu^{(I)}-\mathcal{W}(r_1,\cdots r_n)]/kT},
 \end{equation} 
 such that $\mathcal{B}_0=1$. In his notation $n$ is the number of ions in the pore, $\bar\rho^I$ is the density of permeable ions in bulk  $I$ and $\bar\mu^{(I)}$ is the excess chemical  and electrical potential. As a result, we find in our notation that $\mu^b\sim kT\log(\bar\rho^I)+\bar\mu^{(I)}$. The PMF is continous tracking the positions of all the ions across the whole pore. This differs from our theory which consider the same number of ions in a set of different configurations as ions occupy any binding sites. As a result we can introduce the probability to find $n$ ions inside the pore by summing over all possible configurations that contain $n$ ions, \begin{equation}
     P_n=\sum_{\{n_j\}=n} P(\{n_j\}) 
 \end{equation}and the same is true for the free energy,\begin{equation}
     G_n= \sum_{\{n_j\}=n} G(\{n_j\}). 
 \end{equation} As a result in our theory the binding factor takes the following form, \begin{equation}
     \mathcal{B}_n=e^{(G_0-G_n)/kT}.
 \end{equation} Through comparison of binding factors we are able to recover the following relationship, \begin{equation}
     \sum_{\{n_j\}=n} e^{[G_0-G(\{n_j\})]/kT} = \frac{1}{n!}\int_{c}d\mathbf{r}_1 \cdots \int_c d\mathbf{r}_n e^{[n\mu^b-\mathcal{W}(r_1,\cdots r_n)]/kT}.
 \end{equation}If we cancel the $n!$ terms we are left with,\begin{equation}
     +kT\log\left(\frac{1}{n_0!}\right)+kT\log\left[\sum_{\{n_j\}=n} e^{[(\mathcal{E}(0)-\mathcal{E}(\{n_j\})+\sum_m n_m\Delta \tilde \mu_m]/kT}\right] = +kT\log\left(\int_{c}d\mathbf{r}_1 \cdots \int_c d\mathbf{r}_n e^{[n\mu^b-\mathcal{W}(r_1,\cdots r_n)]/kT}\right).
 \end{equation} where $n_0$ is the number of empty sites with $n$ ions in the pore. In this relationship $\mathcal{E}(\{n_j\})$ represents the total electrostatic energy produced from all the interactions between ions and pore charges when the ions are in the specified configuration $\{n_j\}$.  Finally we note that in the limit that there is only one configuration i.e. if $n=1$ and $M=1$,  then we find,\begin{equation}
     -\Delta \mathcal{E} + \Delta\tilde\mu = 
     +kT\log\left(\int_{c}d\mathbf{r}_1 e^{[\mu^b-\mathcal{W}(r_1)]/kT}\right) 
     \text{ or } 
     -\Delta \mathcal{E} - \bar\mu^c = 
     +kT\log\left(\int_{c}d\mathbf{r}_1 e^{-\mathcal{W}(r_1)/kT}\right).
 \end{equation} In the second equation we have cancelled $\mu^b$ from both sides.

\section{Analysis of Parameters}

The two key parameters within our theory are $\Delta\bar\mu_{Km}$ and $\Delta\bar\mu_{Nam}$, each of which represents the difference in excess chemical potential between the bulk reservoir and site $m$. We derive $\Delta\bar\mu_{Km}$ for sites S1-S4 through fitting of experimentally calculated occupancy's and current-voltage relations. In the main text we have established these parameters for both the wild type (WT) and the mutant (MuT). If we approximate $\bar\mu^b_K$ to be equal to the hydration energy in the bulk $-120kT$ for K$^+$ \cite{marcus1991thermodynamics} we can estimate the value of the potential within the channel and these values are given in table \ref{Tab:MuBarSite}. Krishnapriya et. al. performed free energy calculations the 1K4C (closed) structure  using the AMBER force field to calculate the binding energies at each site (see the paper for details). The authors also found that accounting for quantum mechanical effects only slightly varied the binding energies. The binding energies at S1-S3 were found to be -68, -85 and -117$kT$ respectively and it was found that K$^+$ cannot bind to S4. The latter result has also been observed in \cite{bacstuug2011comparative} who found that S4 cannot be occupied if the two ions are at S1 and S3. It is worth noting that these values strongly vary in contrast to the experimental occupancies which slightly fluctuate around $\sim 0.5$ and occupancy of S4 has been observed \cite{zhou:2004}. Asthagri et. al. \cite{Dixit2009a} calculated the excess chemical potential in S2 using the potential distribution theorem with two potassium ions placed inside the pore. The average excess chemical potential was found to be $\sim -150 kT$ for K$^+$ and $\sim -180kT$ for Na$^+$ although we note that the bulk hydration energies for K$^+$ and Na$^+$ are different and not included here.

Site selectivity is determined within our theory by the difference in excess chemical potential difference, and hence includes the difference in site affinity in addition to the difference in solvation energy within the bulk reservoirs. In \cite{noskov2004control,Noskov2007,egwolf2010ion,kim2011selective} the site selectivities were calculated using molecular dynamics techniques on the closed 1K4C.pdb KcsA channel. Typically three K ions were taken to be in either S1-S4 or possibly in site S0 (a fifth extra-cellular sided site observed in simulation) or in the cavity site close to the pore and a single ion is replace by Na$^+$. The corresponding free energy change is quoted as the selectivity. Although \cite{kim2011selective} also considered the selectivities at various locations within the pore we only consider the values in the oxygen cages that are known to be K$^+$ binding sites. These calculations were performed at differing temperatures and so we convert them into values of kT at 300K.   The values in the table vary significantly both in terms of magnitude but also sign with S4 in \cite{noskov2004control,Noskov2007} being shown to be Na$^+$ selective. As a result we consider the values from \cite{kim2011selective} to calculate $\Delta\bar\mu_{Na,m}$.

\begin{table}[h!]
    \centering
    \begin{tabular}{|c|c|}
\hline 
 Site&  Our theory $\bar\mu^{c,WT}_{mK}$ $[kT]$ \\
\hline 
$m=$1 (S1) & -126.2  \\
\hline 
$m=$2 (S2) & -125.7 \\
\hline 
$m=$3 (S3) & -126  \\
\hline 
$m=$4 (S4) & -126.2 \\
\hline 
\end{tabular} 
\caption{Local interaction at the site in units of $kT$. }\label{Tab:MuBarSite}
\end{table}

\begin{table}[h!]
    \centering
\begin{tabular}{|c|c|c|c|c|c|}
\hline 
Site &   From \cite{noskov2004control,Noskov2007}  [$kT$]  &  From \cite{kim2011selective}  [$kT$]  & From \cite{Dixit2009a} and (\cite{egwolf2010ion}) [$kT$]\\ 
\hline 
$m=$1 (S1)  & +4.2 & +4 &(+7.7)  \\ 
\hline 
$m=$2 (S2)  & +8.5 & +8.3 &+4.5  \\ 
\hline 
$m=$3 (S3)  & +2.9  & +7.6 &N/A  \\ 
\hline 
$m=$4 (S4) & -2  & +6.1 &N/A    \\ 
\hline 
\end{tabular} 
\caption{Selectivities at each site in units $kT$.}\label{Tab:SelectivitySite}
\end{table}

We acknowledge valuable discussions with Bob Eisenberg, Igor Khovanov and Aneta Stefanovska.  The work was funded by a PhD Scholarship from the Faculty of Science and Technology of Lancaster University, the Engineering and Physical Sciences Research Council (grants EP/M016889/1 and EP/M015831/1), and by a Leverhulme Trust Research Project Grant RPG-2017-134.

\bibliographystyle{apsrev}


 \end{document}